\documentclass{ws-ijgmmp}
\usepackage{xcolor}
\usepackage[colorlinks=true,linkcolor=red,citecolor=red,urlcolor=magenta]{hyperref}
\usepackage{orcidlink}
\usepackage{colortbl}
\usepackage{array}
\usepackage[numbers,compress]{natbib}

\makeatletter

\expandafter\let\csname longtable\endcsname\relax
\expandafter\let\csname endlongtable\endcsname\relax
\expandafter\let\csname longtable*\endcsname\relax
\expandafter\let\csname endlongtable*\endcsname\relax

\makeatother

\usepackage{longtable}
\newlength{\gfsep}
\newcommand{\gpanel}[2][]{%
  \setlength{\fboxsep}{\gfsep}\setlength{\fboxrule}{0.4pt}%
  \fcolorbox{black!45}{white}{\includegraphics[#1]{#2}}}
\newenvironment{gridfig}[1]
  {\setlength{\tabcolsep}{0pt}\begin{tabular}{#1}}
  {\end{tabular}}
\AtBeginDocument{%
  \setlength{\abovedisplayskip}{5pt plus 2pt minus 2pt}%
  \setlength{\belowdisplayskip}{5pt plus 2pt minus 2pt}%
  \setlength{\abovedisplayshortskip}{3pt plus 1pt}%
  \setlength{\belowdisplayshortskip}{3pt plus 1pt}%
}
\definecolor{teal}{RGB}{0,128,128}
\numberwithin{equation}{section}
\renewcommand{\theequation}{\arabic{section}.\arabic{equation}}

\newcommand{\hyp}{\,_{2}F_{1}}\newcommand{\dd}{\mathrm{d}}

\makeatletter

\newif\if@fu@BH
\@fu@BHtrue
\newcommand{\BH}{%
  \if@fu@BH black hole~(BH)%
  \global\@fu@BHfalse
  \else BH%
  \fi
}

\newif\if@fu@BHs
\@fu@BHstrue
\newcommand{\BHs}{%
  \if@fu@BHs black holes~(BHs)%
  \global\@fu@BHsfalse
  \else BHs%
  \fi
}

\newif\if@fu@BI
\@fu@BItrue
\newcommand{\BI}{%
  \if@fu@BI Born--Infeld~(BI)%
  \global\@fu@BIfalse
  \else BI%
  \fi
}

\newif\if@fu@AdS
\@fu@AdStrue
\newcommand{\AdS}{%
  \if@fu@AdS anti--de Sitter~(AdS)%
  \global\@fu@AdSfalse
  \else AdS%
  \fi
}

\newif\if@fu@PFDM
\@fu@PFDMtrue
\newcommand{\PFDM}{%
  \if@fu@PFDM perfect fluid dark matter~(PFDM)%
  \global\@fu@PFDMfalse
  \else PFDM%
  \fi
}

\newif\if@fu@GM
\@fu@GMtrue
\newcommand{\GM}{%
  \if@fu@GM global monopole~(GM)%
  \global\@fu@GMfalse
  \else GM%
  \fi
}

\newif\if@fu@CoS
\@fu@CoStrue
\newcommand{\CoS}{%
  \if@fu@CoS cloud of strings~(CoS)%
  \global\@fu@CoSfalse
  \else CoS%
  \fi
}

\newif\if@fu@EH
\@fu@EHtrue
\newcommand{\EH}{%
  \if@fu@EH event horizon~(EH)%
  \global\@fu@EHfalse
  \else EH%
  \fi
}

\newif\if@fu@TD
\@fu@TDtrue
\newcommand{\TD}{%
  \if@fu@TD topological defect~(TD)%
  \global\@fu@TDfalse
  \else TD%
  \fi
}

\newif\if@fu@NLED
\@fu@NLEDtrue
\newcommand{\NLED}{%
  \if@fu@NLED nonlinear electrodynamics~(NLED)%
  \global\@fu@NLEDfalse
  \else NLED%
  \fi
}

\newif\if@fu@ISCO
\@fu@ISCOtrue
\newcommand{\ISCO}{%
  \if@fu@ISCO innermost stable circular orbit~(ISCO)%
  \global\@fu@ISCOfalse
  \else ISCO%
  \fi
}

\newif\if@fu@QPO
\@fu@QPOtrue
\newcommand{\QPO}{%
  \if@fu@QPO quasi-periodic oscillation~(QPO)%
  \global\@fu@QPOfalse
  \else QPO%
  \fi
}

\newif\if@fu@QNM
\@fu@QNMtrue
\newcommand{\QNM}{%
  \if@fu@QNM quasinormal mode~(QNM)%
  \global\@fu@QNMfalse
  \else QNM%
  \fi
}

\newif\if@fu@MCMC
\@fu@MCMCtrue
\newcommand{\MCMC}{%
  \if@fu@MCMC Markov chain Monte Carlo~(MCMC)%
  \global\@fu@MCMCfalse
  \else MCMC%
  \fi
}

\newif\if@fu@VdW
\@fu@VdWtrue
\newcommand{\VdW}{%
  \if@fu@VdW Van der Waals~(VdW)%
  \global\@fu@VdWfalse
  \else VdW%
  \fi
}

\newif\if@fu@GBF
\@fu@GBFtrue
\newcommand{\GBF}{%
  \if@fu@GBF
    greybody factor~(GBF)%
    \global\@fu@GBFfalse
  \else
    GBF%
  \fi
}

\newif\if@fu@CS
\@fu@CStrue
\newcommand{\CS}{%
  \if@fu@CS
    cloud of strings~(CoS)%
    \global\@fu@CSfalse
  \else
    CoS%
  \fi
}

\newif\if@fu@GBFs
\@fu@GBFstrue
\newcommand{\GBFs}{%
  \if@fu@GBFs
    greybody factors~(GBFs)%
    \global\@fu@GBfsfalse
  \else
    GBFs%
  \fi
}

\makeatother

\begin{document}

\markboth{F. Ahmed et al.}
{Imprints of PFDM distributions and topological defects on Born--Infeld AdS-BH signatures}

\catchline{}{}{}{}{}

\title{Imprints of perfect fluid dark matter distributions and topological defects on Born--Infeld AdS black hole signatures
}

\author{Faizuddin Ahmed\orcidlink{0000-0003-2196-9622}
}
\address{Department of Physics, The Assam Royal Global University, Guwahati, 781035, Assam, India\\
\email{faizuddinahmed15@gmail.com}
}

\author{Ahmad Al-Badawi\orcidlink{0000-0002-3127-3453}
}
\address{Department of Physics, Al-Hussein Bin Talal University, 71111, Ma'an, Jordan
\email{ahmadbadawi@ahu.edu.jo}
}

\author{\.{I}zzet Sakall{\i}\orcidlink{0000-0001-7827-9476}
}
\address{Physics Department, Eastern Mediterranean University, Famagusta 99628, North Cyprus via Mersin 10, Turkey
\email{izzet.sakalli@emu.edu.tr}
}

\maketitle

\begin{abstract}
We construct a static, spherically symmetric \BI{} \AdS{} \BH{} that carries a topological defect, either a \GM{} or a \CS{}, and is immersed in \PFDM{}. The metric function collects four independent deformations of the Schwarzschild geometry: a constant solid-angle deficit $(1-\xi)$ from the defect, the nonlinear charge sector controlled by the \BI{} scale $\beta$, the logarithmic \PFDM{} term set by the intensity $\lambda$, and the \AdS{} curvature $\Lambda$. We verify that the solution reduces to the charged \AdS{}--\PFDM{} defect geometry as $\beta \to \infty$ and to Schwarzschild when all deformations switch off. The horizon, photon sphere, and shadow all grow monotonically with the defect strength $\xi$. We compute the innermost stable circular orbit and the epicyclic frequencies, read off the relativistic-precession predictions for twin-peak quasi-periodic oscillations, and confront them with five accreting sources through a Markov chain Monte Carlo fit. For spin-0 perturbations we obtain the scalar potential, the higher-order WKB \QNM{} spectrum, and the \GBF{}, all of which soften as $\xi$ rises. The extended phase space analysis yields a \VdW{}-type $P$--$V$ structure whose critical point shifts with the defect, the \BI{} scale, and the halo intensity. Throughout, the four sectors leave distinct imprints on imaging, timing, ringdown, and thermal channels, {\color{black} which we read as parameter sensitivities rather than as an observational separation of the four, a distinction a joint multichannel fit would be needed to settle}.
\end{abstract}


\section{Introduction}\label{isec1}
Astrophysical \BHs{} are not isolated objects. They sit at the centres of galaxies,
inside dark matter haloes, and are fed by accreting plasma, so the geometry they
produce departs from the vacuum Schwarzschild and Kerr solutions. For a long time
that departure was a formal question. It has become a measurable one. The Event
Horizon Telescope has resolved the emission rings of M87$^{\ast}$ and
Sgr~A$^{\ast}$~\cite{Akiyama:2019, Akiyama:2022}, and twin-peak quasi-periodic
oscillations from accreting binaries sample orbital timescales a few gravitational
radii outside the horizon~\cite{IngramMotta:2019}. The question these data raise is
which modifications of the near-horizon metric they can tell apart, and which they
cannot. We take up that question here for one spacetime.

The spacetime is a static, spherically symmetric \BH{} whose metric function carries four
deformations of the Schwarzschild geometry. Each has its own physical origin and its
own parameter. We name all four here, so that the rest of the paper can be read
against a fixed picture. The first is the dark matter halo in which the hole is
embedded; modelled as a perfect fluid, it enters the metric function through a logarithmic term
of intensity $\lambda$. The second is the electromagnetic sector, which obeys \BI{}
rather than Maxwell electrodynamics and so removes the divergent self-energy of a
point charge, at the cost of a new scale $\beta$. The third is a topological defect,
either a \GM{} or a \CS{}. Whichever it is, it removes a fixed fraction of the solid
angle, so the metric function tends to $(1-\xi)$ instead of unity far from the hole. The fourth
is a negative cosmological constant $\Lambda$, which makes the spacetime
asymptotically \AdS{} and lets the thermodynamics be posed in an extended phase space
with $\Lambda$ acting as a pressure. These four are independent. Switching any one off
leaves the other three untouched, and switching off all four returns Schwarzschild.

The halo term has the longest observational pedigree of the four. Rotation curves,
gravitational lensing and the cosmic microwave background all point to a dark
component that dominates the matter budget and collects around essentially every
galaxy~\cite{Rubin:1970, Bertone:2018}. \PFDM{} is a tractable analytic model of it,
threading the halo into the metric through a logarithmic
potential~\cite{Kiselev:2002dx, Kiselev2003, LiYang2012}. It has since been built into
charged, rotating and modified-gravity backgrounds~\cite{Xu:2014, Xu:2018}. The appeal
is practical. One intensity parameter carries the halo while the field equations stay
solvable, so its imprint can be followed analytically from one observational channel
to the next.

The optical channel is the best developed of those. Horizon-scale imaging turned the
photon sphere and the shadow into measured quantities, and a large literature now maps
how matter and additional fields reshape the shadow and the bending of
light~\cite{Mirekhtiary:2023}. Strong-field lensing has been worked out for a range of
backgrounds, for an observer at infinity and at finite
distance~\cite{Gursel:2021, HiokiMiyamoto:2025}, including geometries carrying a
topological defect~\cite{LuGM:2025}, and shadow calculations for \BHs{} in dark matter
or exotic fields continue to appear~\cite{Uniyal:2023}. The optical appearance is by
now a quantitative probe of the near-horizon metric.

The remaining two deformations are theoretical in origin rather than observational. 
\BI{} theory was introduced to cure the divergent self-energy of a point charge and returns Maxwell's theory as
$\beta\to\infty$~\cite{Born:1934gh, Gunasekaran:2012dq}; it is now standard in regular
and charged \BH{} solutions and in their lensing and thermodynamic
response~\cite{Mazharimousavi:2009}. Numerous authors have studied black hole solutions with a cosmological constant in Einstein-Born-Infeld gravity, for instances, charged black hole \cite{Fernando2003,BelmarHerrera2022}; asymptotically black hole solutions in arbitrary dimension \cite{Dey2004,Cai2004} with their critical behaviors \cite{Zou2014,Banerjee2012,Miskovic2008}; Ruppeiner geometry \cite{Chen2023,Kumara2021}; Lyapunov exponent and phase transition  \cite{Yang2023}; gravitational lensing \cite{Aydiner2025}; and as a heat engine \cite{Johnson2016} were investigated. A \CS{}~\cite{Letelier:1979ej, Goto1971} and a
\GM{}~\cite{Barriola:1989hx} come from different microphysics, yet they act on a
static spherically symmetric metric in the same way, by removing a constant fraction
of the solid angle. That shared action is useful here, because a single calculation in
$\xi$ then covers both cases, with $\xi=\alpha$ for the string cloud and
$\xi=8\pi\eta^{2}$ for the monopole.

The \AdS{} sector is what makes the thermodynamics nontrivial. Treating $\Lambda$ as a
pressure $P=-\Lambda/8\pi$ promotes it to an extended phase space in which the mass
plays the role of enthalpy~\cite{Kastor:2009wy, Dolan:2010ha, Dolan:2011xt}. Charged
\AdS{} \BHs{} then show a \VdW{} liquid--gas transition with a sharp critical
point~\cite{Chamblin:1998, Kubiznak:2012wp}, and the same structure has been mapped
for Born--Infeld, Gauss--Bonnet and dark matter corrected backgrounds, where reentrant
transitions and richer critical behavior appear~\cite{Cai:2014, Zou:2014, Wu:2020}.
Holographic and microstructure readings of this chemistry have
followed~\cite{Yerra:2021, Yerra:2022}, alongside work folding generalized entropies
into the first law~\cite{Pourhassan:2025}. The \AdS{} curvature is a control parameter
of the phase behavior, not a passive backdrop.

Two dynamical probes complete the set we use. Perturbations of the hole ring down at
quasinormal frequencies, which the WKB method computes to high
order~\cite{KonoplyaZhidenkoZinhailo:2019}, and the same potential barrier fixes the
\GBF{} that filters the Hawking flux. Twin-peak quasi-periodic oscillations, already
mentioned above, are tied to the orbital and epicyclic frequencies by the
relativistic-precession and resonance models~\cite{StellaVietri:1998,
AbramowiczKluzniak:2001, Wagoner:2001, Stuchlik:2013}. We have applied probes of this
kind to a range of modified backgrounds~\cite{Sakalli:2015, FA1,FA2, Sakalli:2022,
Sucu:2026,FA5}. What has not been done, so far as we are aware, is to put all four
deformations in one metric and follow them together through imaging, timing, ringdown
and thermodynamics. Pairs of them have been studied often enough. The joint problem is
the one that decides whether the parameters can be separated observationally or only
ever measured in combination, and that is the gap we address.

The paper is organized as follows. Section~\ref{isec2} derives the metric from the
field equations, verifies its limits, and locates the horizon. Section~\ref{isec3}
treats null geodesics, giving the photon sphere, the shadow radius seen by an observer
at finite distance, the ray trajectories, and the eikonal link between the photon orbit
and the ringdown. Section~\ref{isec4} turns to timelike orbits, the innermost stable
circular orbit, the periastron advance and the epicyclic frequencies, which
Section~\ref{isec5} confronts with twin-peak data from five accreting sources through a
Markov chain Monte Carlo fit. Section~\ref{isec6} computes the scalar quasinormal
spectrum and the greybody factor, Section~\ref{isec7} develops the extended phase space
thermodynamics and the Hawking emission rate, and Section~\ref{isec8} concludes.

\section{BI--AdS BH with PFDM and a topological defect}\label{isec2}

We now derive the background geometry announced in Sec.~\ref{isec1}. The action is
\begin{align}
\mathcal{S}=\int \dd^{4}x\sqrt{-g}\left[\frac{R-2\Lambda+\mathcal{L}(\mathcal{F})}{2\kappa}+\mathcal{L}_{\rm DM}+\mathcal{L}_{\rm TD}\right],\label{aa1}
\end{align}
where $R$ is the Ricci scalar, $\kappa=8\pi$, $\Lambda$ is the cosmological constant, and $g$ is the determinant of the metric $g_{\mu\nu}$. The term $\mathcal{L}_{\rm DM}$ is the Lagrangian density of the dark matter field~\cite{Kiselev2003, LiYang2012}, and $\mathcal{L}_{\rm TD}$ that of the topological defect, a \GM{} or a \CS{}~\cite{Letelier:1979ej, Barriola:1989hx}. The \BI{} sector reads
\begin{align}
\mathcal{L}(\mathcal{F})=4\beta^{2}\left(1-\sqrt{1+\frac{\mathcal{F}}{2\beta^{2}}}\right),
\end{align}
with $\beta$ the \BI{} scale, $\mathcal{F}=F_{\mu\nu}F^{\mu\nu}$ the field invariant, and $F_{\mu\nu}=\partial_{\mu}A_{\nu}-\partial_{\nu}A_{\mu}$ built from the four-potential $A_{\mu}$. Maxwell electrodynamics returns as $\beta\to\infty$; finite $\beta$ caps the field strength and regularizes the self-energy of the charge.

A \GM{} is a triplet of scalar fields whose global $O(3)$ symmetry breaks spontaneously to $U(1)$, and whose gravitational field produces a solid-angle deficit. Its matter sector is~\cite{Barriola:1989hx}
\begin{equation}
\mathcal{L}^{\rm GM}_{m}=-\frac{1}{2}\partial_{\mu}\psi^{a}\partial^{\mu}\psi^{a}-\frac{\chi}{4}\left(\psi^{a}\psi^{a}-\eta^{2}\right)^{2},
\end{equation}
with $a=1,2,3$ labelling the components, $\chi$ the self-coupling, and $\eta$ the symmetry-breaking scale. We take the hedgehog ansatz
\begin{equation}
\psi^{a}=\eta\,\frac{x^{a}}{r},\qquad x^{a}x^{a}=r^{2},
\end{equation}
which holds outside the monopole core.

To include string-like objects instead, we use the Nambu--Goto action~\cite{Letelier:1979ej}
\begin{equation}
\mathcal{S}_{\rm CS}=\int \mathcal{M}\sqrt{-\gamma}\,\dd\zeta^{0}\,\dd\zeta^{1}
=\int \mathcal{M}\sqrt{-\tfrac{1}{2}\Sigma^{\mu\nu}\Sigma_{\mu\nu}}\,\dd\zeta^{0}\,\dd\zeta^{1},
\label{action-4}
\end{equation}
where $\mathcal{M}$ is the dimensionless string constant and $(\zeta^{0},\zeta^{1})$ the timelike and spacelike world-sheet parameters~\cite{Goto1971}. We write these as $\zeta^{a}$ rather than $\lambda^{a}$ to keep them clear of the \PFDM{} intensity $\lambda$ below. The induced-metric determinant and the world-sheet bi-vector read
\begin{equation}
\gamma=g_{\mu\nu}\frac{\partial x^{\mu}}{\partial\zeta^{a}}\frac{\partial x^{\nu}}{\partial\zeta^{b}},\qquad
\Sigma^{\mu\nu}=\epsilon^{ab}\frac{\partial x^{\mu}}{\partial\zeta^{a}}\frac{\partial x^{\nu}}{\partial\zeta^{b}},
\end{equation}
with $\epsilon^{01}=-\epsilon^{10}=1$, and the energy--momentum tensor is
\begin{equation}
T_{\mu\nu}^{\rm CS}
=2\frac{\partial}{\partial g_{\mu\nu}}\left(\mathcal{M}\sqrt{-\tfrac{1}{2}\Sigma^{\mu\nu}\Sigma_{\mu\nu}}\right)
=\frac{\rho^{\rm CS}\,\Sigma_{\alpha\nu}\,\Sigma_{\mu}{}^{\alpha}}{\sqrt{-\gamma}},
\label{action-5}
\end{equation}
where $\rho^{\rm CS}$ is the proper density of the cloud.

Variation of the action~\eqref{aa1} gives the gravitational and electromagnetic field equations
\begin{align}
G_{\mu}{}^{\nu}-\tfrac{1}{2}g_{\mu}{}^{\nu}\mathcal{L}(\mathcal{F})+2\mathcal{L}_{\mathcal{F}}F_{\mu\sigma}F^{\sigma\nu}&=8\pi\,T_{\mu}{}^{\nu},\\
\nabla_{\mu}\!\left(\sqrt{-g}\,\mathcal{L}_{\mathcal{F}}F^{\mu\nu}\right)&=0,
\end{align}
with $\mathcal{L}_{\mathcal{F}}=\dd\mathcal{L}/\dd\mathcal{F}$, $G_{\mu}{}^{\nu}=R_{\mu}{}^{\nu}-\tfrac{1}{2}g_{\mu}{}^{\nu}R$, and a total matter stress-energy tensor $T_{\mu}{}^{\nu}=T_{\mu}{}^{\nu}(\mathrm{DM})+T_{\mu}{}^{\nu}(\mathrm{TD})$ built from the dark matter and topological-defect contributions. The \PFDM{} ~\cite{Kiselev2003, LiYang2012} and defect contributions are
\begin{align}
T_{\mu}{}^{\nu}(\mathrm{DM})&=\mathrm{diag}\!\left(\frac{\lambda}{\kappa r^{3}},\,\frac{\lambda}{\kappa r^{3}},\,-\frac{\lambda}{2\kappa r^{3}},\,-\frac{\lambda}{2\kappa r^{3}}\right),\\
T_{\mu}{}^{\nu}(\mathrm{TD})&=\mathrm{diag}\!\left(-\frac{\xi}{r^{2}},\,-\frac{\xi}{r^{2}},\,0,\,0\right),
\end{align}
where $\xi=\alpha$ for a \CS{}~\cite{Letelier:1979ej} and $\xi=\kappa\eta^{2}$ for a \GM{}~\cite{Barriola:1989hx}, with $\eta$ the symmetry-breaking scale. The defect therefore acts through a single constant $\xi$, whatever its microscopic origin.

For the static, spherically symmetric line element
\begin{equation}
\dd s^{2}=-f(r)\,\dd t^{2}+\frac{\dd r^{2}}{f(r)}+r^{2}\big(\dd\theta^{2}+\sin^{2}\theta\,\dd\phi^{2}\big),
\label{metric}
\end{equation}
with \BI{} electrodynamics, a \PFDM{} fluid of intensity $\lambda$, and a defect contributing $(1-\xi)$, the field equations give the metric function
\begin{align}
f(r)&=1-\xi-\frac{2M}{r}+\frac{2\beta^{2}r^{2}}{3}\!\left(1-\sqrt{1+\frac{Q^{2}}{\beta^{2}r^{4}}}\right)+\frac{4Q^{2}}{3r^{2}}\,\hyp\!\left(\tfrac14,\tfrac12,\tfrac54,-\frac{Q^{2}}{\beta^{2}r^{4}}\right)\nonumber\\
&+\frac{\lambda}{r}\ln\!\frac{r}{|\lambda|}-\frac{\Lambda r^{2}}{3}.\label{lapse}
\end{align}
The four deformations are independent: $\xi$ fixes the asymptotic value, $\beta$ sets the nonlinear charge scale, $\lambda$ the halo intensity, and $\Lambda$ the curvature. {\color{black} Note that in the limiting case $\xi \to 0$, one recovers the Born--Infeld AdS black holes surrounded by \PFDM{} recently reported in \cite{Panah2026}.} {\color{black} A word on the parameters is in order. In geometric units the deficit $\xi$ is dimensionless, and physical defects keep it small, $0\le\xi\ll1$; we carry $\xi$ toward unity only to display the trends. The \PFDM{} intensity $\lambda$ carries the dimension of length and sets the halo scale, while the \BI{} parameter $\beta$ has the dimension of inverse length and caps the electric field, with the Maxwell limit at $\beta\to\infty$. Because the defect enters the geometry only through the constant $(1-\xi)$, a cloud of strings with $\xi=\alpha$ and a global monopole with $\xi=8\pi\eta^{2}$ produce identical exterior spacetimes, so an observation that fixes $\xi$ pins down the deficit but cannot by itself separate a string cloud from a monopole; telling the two apart needs independent information about the field content.}

\begin{figure}[ht!]
\centering
\begin{gridfig}{@{}c@{\hspace{3pt}}c@{}}
\gpanel[width=0.47\textwidth]{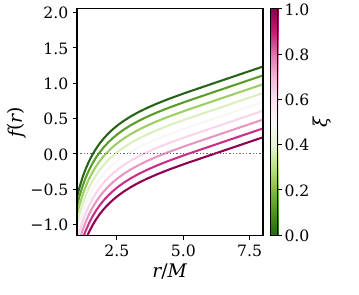} &
\gpanel[width=0.47\textwidth]{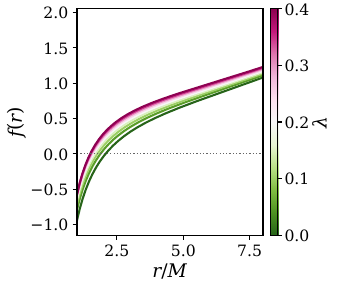}\\[2pt]
(a) & (b)
\end{gridfig}
\caption{\label{fig:metric}Lapse $f(r)$ for the \BI{}--\AdS{} \BH{} with \PFDM{} and a topological defect. Left: fixed $Q=0.4$, $\beta=1$, $\lambda=0.1$, $\Lambda=-0.02$, varying the defect strength $\xi\in[0,1]$. Right: fixed $\xi=0.1$, varying the halo intensity $\lambda$. The dotted line marks $f=0$.}
\end{figure}

Four limits anchor the solution, each confirmed by symbolic computation. As $\beta \to \infty$ the nonlinear sector collapses to the Maxwell term,
\begin{equation}
f(r)\to 1-\xi-\frac{2M}{r}+\frac{Q^{2}}{r^{2}}+\frac{\lambda}{r}\ln\!\frac{r}{|\lambda|}-\frac{\Lambda r^{2}}{3},
\end{equation}
the charged \AdS{} geometry with \PFDM{} and a defect. In the opposite, strong-coupling limit $\beta\to0$ the charge contribution drops out altogether, because \BI{} theory caps the field and leaves only the non-electromagnetic sectors. {\color{black} Moreover, in the limit $\beta  \to \infty$ and $\lambda \to 0$, we get back the space-time around a global monopole swallowed by a charged AdS black hole \cite{Guendelman1991}.} Switching off all four deformations, $\xi=\lambda=\Lambda=Q=0$, returns the Schwarzschild value $f=1-2M/r$; and for $\Lambda=0$ the metric function tends to $1-\xi$ at large $r$, the deficit expected from the defect. Figure~\ref{fig:metric} plots the metric function~\eqref{lapse} and shows a single event horizon, which moves outward as the defect strength $\xi$ grows and inward as the halo intensity $\lambda$ grows.

\subsection{Horizon structure}\label{isec2a}
The deficit, the halo, and the \AdS{} term each shift the zero-crossing of the metric function, so we read the horizon count and location off numerically. The \EH{}s are the positive roots of $f(r_{h})=0$, and since Eq.~\eqref{lapse} mixes a hypergeometric charge sector with a logarithmic halo term we locate them by a sign-change search at resolution $\Delta r=0.01M$. For the representative set $Q=0.4$, $\beta=1$, $\lambda=0.1$, $\Lambda=-0.02$, a stronger defect pushes the outer \EH{} from $r_{h}=1.596M$ at $\xi=0$ out to $2.893M$ at $\xi=0.5$ (Table~\ref{tab:horizon}). The mechanism is simple. The deficit lowers $f(r)$ at fixed $r$, so the zero-crossing migrates outward, and the same shift enlarges the photon sphere and shadow of Sec.~\ref{isec3}. {\color{black} The horizon count depends on the sign of the curvature term. In the \AdS{} branch used throughout the body ($\Lambda<0$) the metric function rises without bound at large $r$, so $f(r_{h})=0$ carries at most two positive roots, an inner Cauchy horizon and an outer \EH{}. When the curvature term is relaxed toward and past $\Lambda=0$, the \BI{} charge and the \PFDM{} logarithm can add a third crossing, and the geometry then admits up to three positive horizons, an inner Cauchy horizon, an outer \EH{}, and a cosmological horizon, reproducing the horizon count reported for the defect-free \BI{}--\PFDM{} solution~\cite{Panah2026}. For $M=0.6$, $Q=0.4$, $\beta=2$, $\lambda=-0.3$, $\Lambda=0.02$ the horizons sit at $r_{h}=0.088M$, $1.642M$, and $10.902M$ at $\xi=0$; raising the defect to $\xi=0.2$ holds the inner horizon almost fixed, drives the \EH{} outward to $2.267M$, and pulls the cosmological horizon inward to $9.156M$, so the defect compresses the static region between the event and cosmological horizons.}

\begin{table}[ht]
\centering
\small
\caption{Outer \EH{} radius $r_h$, photon-sphere radius $r_{\rm ph}$,
and shadow radius $R_{\rm sh}$ (in units of $M$) for
$Q=0.4$, $\beta=1$, $\lambda=0.1$, and $\Lambda=-0.02$ as the defect
strength $\xi$ varies.}
\label{tab:horizon}
\resizebox{\ifdim\width>\columnwidth\columnwidth\else\width\fi}{!}{%
\begin{tabular}{|c|c|c|c|}
\hline
\textbf{$\xi$} & \textbf{$r_h/M$} & \textbf{$r_{\rm ph}/M$} &
\textbf{$R_{\rm sh}/M$} \\
\hline
0.00 & 1.5958 & 2.4397 & 3.9696 \\

0.10 & 1.7621 & 2.7078 & 4.5516 \\

0.20 & 1.9630 & 3.0407 & 5.2653 \\

0.30 & 2.2089 & 3.4655 & 6.1399 \\

0.40 & 2.5135 & 4.0270 & 7.1968 \\

0.50 & 2.8934 & 4.8051 & 8.4259 \\
\hline
\end{tabular}
}
\end{table}

\section{Photon sphere, shadow, and null trajectories}\label{isec3}

Null geodesics fix the optical appearance of the \BH{}, so the photon sphere, the
shadow, and the bending of individual rays translate the four metric deformations
into quantities that horizon-scale imaging can reach.

\subsection{Null geodesics and the effective potential}\label{isec3a}
\begin{figure}[ht!]
\centering
\begin{gridfig}{@{}c@{\hspace{3pt}}c@{}}
\gpanel[width=172.1pt,height=102.4pt]{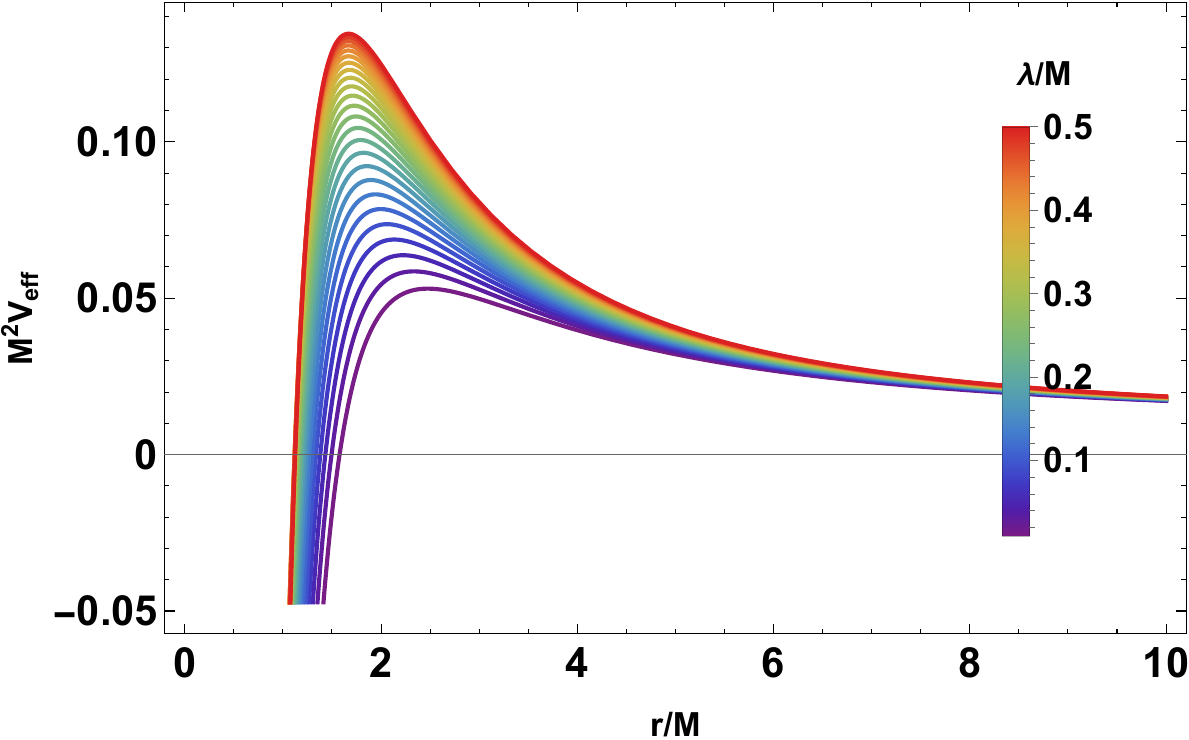} &
\gpanel[width=172.1pt,height=102.4pt]{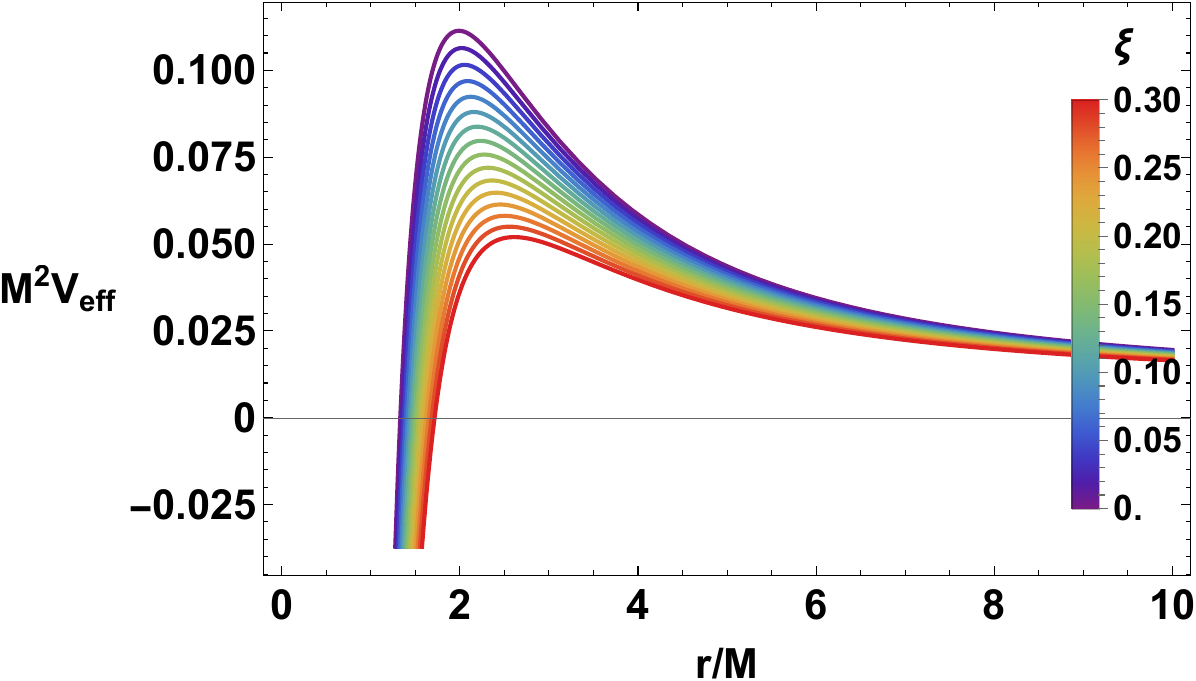}\\[2pt]
(i) $\xi=0.1$ & (ii) $\lambda/M=0.5$
\end{gridfig}
\caption{\label{fig:potential-null}Null effective potential $V_{\rm n}(r)=L^{2}f(r)/r^{2}$
versus radius. Left: fixed $\xi=0.1$, varying the halo intensity $\lambda$. Right:
fixed $\lambda/M=0.5$, varying the defect strength $\xi$. The remaining parameters
are $Q/M=0.5$, $\beta=0.1/M$, $\Lambda=-0.03$, $L=1$. The barrier peak, whose
location is the photon sphere, drops and moves outward as either parameter grows.}
\end{figure}

A photon in the equatorial plane ($\theta=\pi/2$, $\dot\theta=0$) of the
metric~\eqref{metric} follows $\dd s^{2}=0$, which reads
\begin{equation}
-f(r)\,\dot t^{2}+\frac{\dot r^{2}}{f(r)}+r^{2}\dot\phi^{2}=0.\label{n1}
\end{equation}
Staticity and axial symmetry supply two constants of motion,
\begin{equation}
E=f(r)\,\dot t,\qquad L=r^{2}\dot\phi,\label{n2}
\end{equation}
the conserved energy and angular momentum of the ray. Substituting Eq.~\eqref{n2}
into Eq.~\eqref{n1} collapses the motion to a one-dimensional radial problem,
\begin{equation}
\dot r^{2}+V_{\rm n}(r)=E^{2},\qquad V_{\rm n}(r)=\frac{L^{2}}{r^{2}}\,f(r),\label{n3}
\end{equation}
with $V_{\rm n}$ the null effective potential. Figure~\ref{fig:potential-null}
shows it for a family of halo intensities and a family of defect strengths. Both
deformations lower and broaden the barrier, because either one weakens the
centrifugal wall that turns photons back. That is the first sign that both push the
photon sphere outward.

We follow the trajectory in the orbital variable $u=1/r$. Dividing the radial
equation by $\dot\phi^{2}=L^{2}u^{4}$ and writing $b\equiv L/E$ for the impact
parameter gives
\begin{equation}
\left(\frac{\dd u}{\dd\phi}\right)^{2}=\frac{1}{b^{2}}-f\!\left(\tfrac1u\right)u^{2},
\label{n4}
\end{equation}
and one differentiation removes the constant,
\begin{equation}
\frac{\dd^{2}u}{\dd\phi^{2}}=\frac{1}{2}\,f'\!\left(\tfrac1u\right)-u\,f\!\left(\tfrac1u\right).
\label{n5}
\end{equation}
Equation~\eqref{n5} is what we integrate for the ray bundles below. Since $b$ enters only
through the turning point of Eq.~\eqref{n4}, one solve covers every ray once the starting
data are fixed.

\subsection{Photon sphere and shadow radius}\label{isec3b}
\begin{figure}[ht!]
\centering
\begin{gridfig}{@{}c@{\hspace{3pt}}c@{}}
\gpanel[width=172.1pt,height=120.4pt]{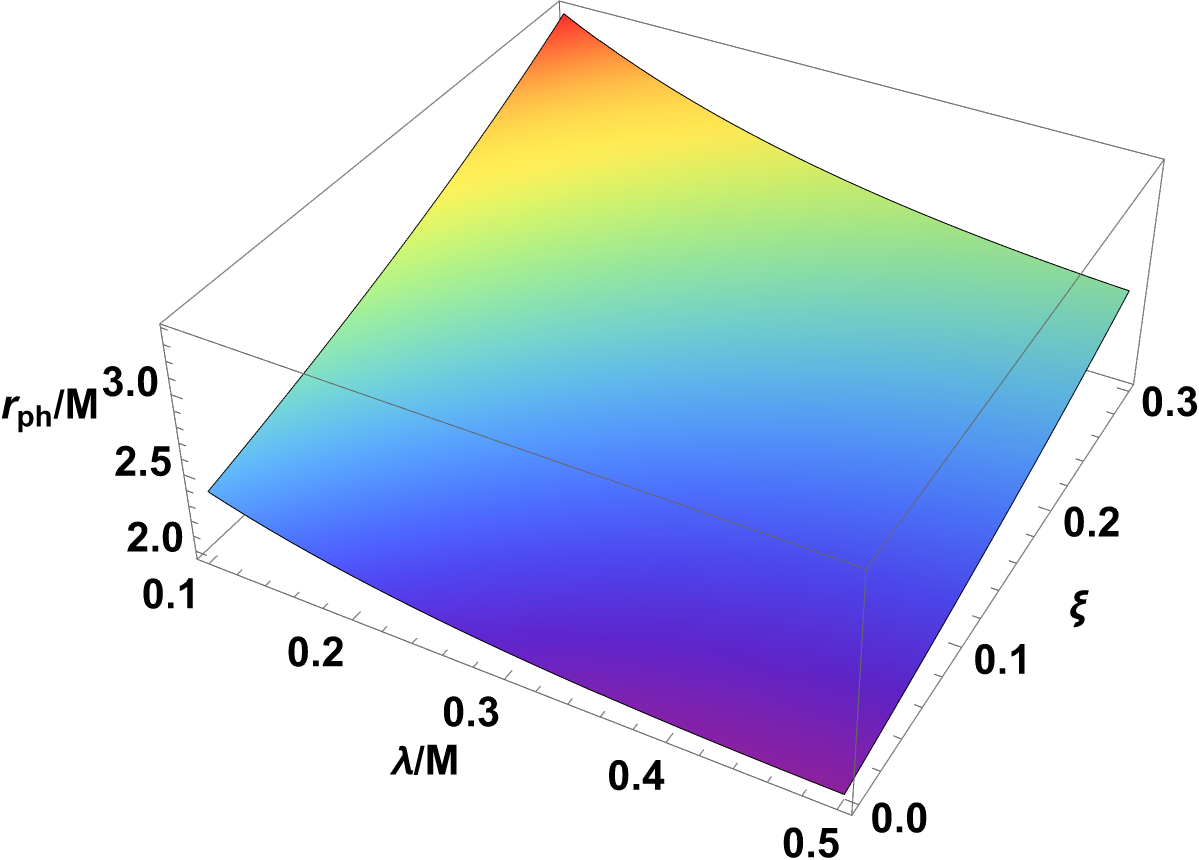} &
\gpanel[width=172.1pt,height=120.4pt]{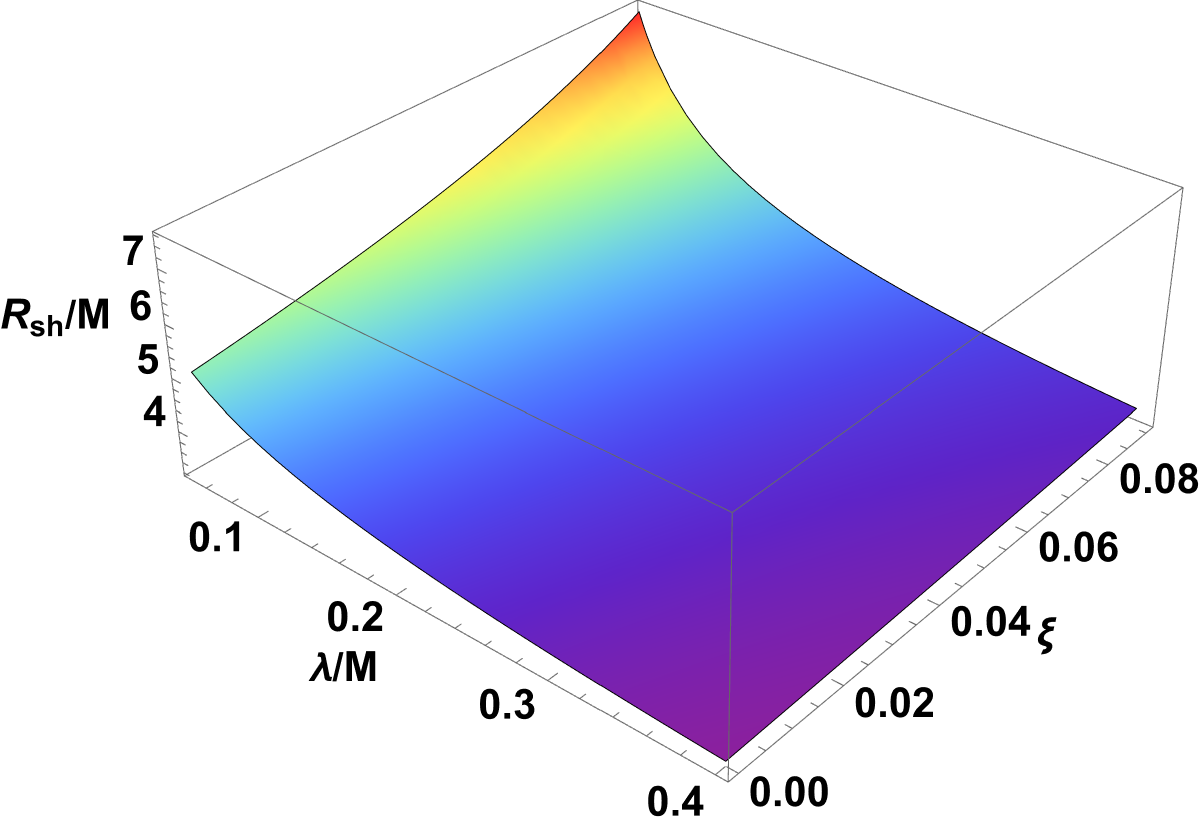}\\[2pt]
(a) & (b)
\end{gridfig}
\caption{\label{fig:photon3d}Photon sphere and shadow surfaces at $Q/M=0.5$,
$\beta=0.2/M$. (a) Photon sphere radius $r_{\rm ph}(\lambda/M,\xi)$. (b)
Distant-observer shadow radius $R_{\rm sh}(\lambda/M,\xi)$ from Eq.~\eqref{shadow-3}.
Both surfaces rise monotonically with the defect strength $\xi$ and fall with the
halo intensity $\lambda$; the shadow tracks the photon sphere reduced by the
$\sqrt{1-\xi}$ factor.\label{fig:shadow3d}}
\end{figure}

A circular photon orbit sits where the null potential is stationary,
$V_{\rm n}'(r_{\rm ph})=0$, equivalently
\begin{equation}
2f(r_{\rm ph})-r_{\rm ph}\,f'(r_{\rm ph})=0.\label{n6}
\end{equation}
Carried out on the metric function~\eqref{lapse}, this condition reads
\begin{align}
1-\xi-\frac{3M}{r}+\frac{\lambda}{2r}\!\left(3\ln\frac{r}{|\lambda|}-1\right)
-\frac{2Q^{2}}{3r^{2}\sqrt{1+\dfrac{Q^{2}}{\beta^{2}r^{4}}}}
&+\frac{8Q^{2}}{3r^{2}}\,\hyp\!\left(\tfrac14,\tfrac12;\tfrac54;-\tfrac{Q^{2}}{\beta^{2}r^{4}}\right)\nonumber\\
&-\frac{4Q^{4}}{15\beta^{2}r^{6}}\,\hyp\!\left(\tfrac54,\tfrac32;\tfrac94;-\tfrac{Q^{2}}{\beta^{2}r^{4}}\right)=0.
\label{n7}
\end{align}
The hypergeometric terms block a closed form, so $r_{\rm ph}$ follows numerically. The
orbit is unstable, $V_{\rm n}''(r_{\rm ph})<0$, so a photon nudged off it either spirals
to the horizon or escapes; that instability is what makes the photon sphere the edge of
the shadow. The critical value the impact parameter reaches there,
\begin{equation}
b_{\rm c}=\frac{r_{\rm ph}}{\sqrt{f(r_{\rm ph})}},\label{n8}
\end{equation}
marks the rim of the dark region for an observer where the metric function is asymptotically constant.
This spacetime is not asymptotically flat: with $\Lambda\neq0$, and already through the
deficit at $\Lambda=0$, $\lim_{r\to\infty}f(r)\neq1$, so the apparent size must be referred
to a static observer at finite radius $r_{0}$. Following Perlick and
Tsupko~\cite{Perlick2022}, that observer measures
\begin{equation}
R_{\rm sh}=r_{\rm ph}\sqrt{\frac{f(r_{0})}{f(r_{\rm ph})}}.\label{shadow-1}
\end{equation}
Inserting the metric function~\eqref{lapse} at $\Lambda=0$ and sending the observer to large
radius, where $f(r_{0})\to1-\xi$, reduces this to {\small
\begin{equation}
R_{\rm sh}=\frac{\sqrt{1-\xi}\,r_{\rm ph}}
{\sqrt{\,1-\xi-\dfrac{2M}{r_{\rm ph}}
+\dfrac{2\beta^{2}r_{\rm ph}^{2}}{3}\!\left(1-\sqrt{1+\dfrac{Q^{2}}{\beta^{2}r_{\rm ph}^{4}}}\right)
+\dfrac{4Q^{2}}{3r_{\rm ph}^{2}}\,\hyp\!\left[\tfrac14,\tfrac12,\tfrac54,-\dfrac{Q^{2}}{\beta^{2}r_{\rm ph}^{4}}\right]
+\dfrac{\lambda}{r_{\rm ph}}\ln\dfrac{r_{\rm ph}}{|\lambda|}}}\,,\label{shadow-3}
\end{equation}
}
which is $b_{\rm c}$ of Eq.~\eqref{n8} scaled by the deficit factor $\sqrt{1-\xi}$. The
defect therefore enters the apparent shadow twice, once through $r_{\rm ph}$ and
$f(r_{\rm ph})$ and once through the asymptotic rescaling, and the same double dependence
governs finite-distance lensing by a \GM{}~\cite{LuGM:2025}. Tables~\ref{tab:horizon}
and~\ref{tab:shadowobs} report $b_{\rm c}$ at the small $|\Lambda|$ used there, where it
is an accurate proxy for the angular size, while Table~\ref{tab:photon-shadow-lam}
applies the distant-observer expression~\eqref{shadow-3} with the $\sqrt{1-\xi}$ factor
included, at fixed $Q/M=0.5$ and $\beta=0.2/M$. The photon sphere shrinks with $\lambda$
and grows with $\xi$, and the two trends compete: a heavier halo pulls the orbit inward
through the logarithmic well, while a stronger defect pushes it outward through the
deficit. Figure~\ref{fig:photon3d} renders both surfaces, with the rise along the defect
axis the steeper of the two slopes in panel (a) and the matching shadow in panel (b).

\begin{table}[ht]
\centering
\small
\setlength{\tabcolsep}{4pt}
\caption{Photon-sphere radius $r_{\rm ph}/M$ from Eq.~\eqref{n7} and
distant-observer shadow radius $R_{\rm sh}/M$ from Eq.~\eqref{shadow-3}
for the halo intensities $\lambda$ and defect strengths $\xi$ indicated, at
$Q/M=0.5$ and $\beta=0.2/M$. The two halves of the table continue one another.}
\label{tab:photon-shadow-lam}
\resizebox{\ifdim\width>\columnwidth\columnwidth\else\width\fi}{!}{%
\begin{tabular}{|c|c|c|c||c|c|c|c|}
\hline
\textbf{$\lambda/M$} & \textbf{$\xi$} & \textbf{$r_{\rm ph}/M$} & \textbf{$R_{\rm sh}/M$} &
\textbf{$\lambda/M$} & \textbf{$\xi$} & \textbf{$r_{\rm ph}/M$} & \textbf{$R_{\rm sh}/M$}\\
\hline
0.05 & 0.00 & 2.54 & 4.43 & 0.25 & 0.00 & 2.10 & 3.52\\
0.05 & 0.10 & 2.83 & 4.94 & 0.25 & 0.10 & 2.31 & 3.87\\
0.05 & 0.20 & 3.20 & 5.57 & 0.25 & 0.20 & 2.57 & 4.30\\
0.05 & 0.30 & 3.67 & 6.37 & 0.25 & 0.30 & 2.91 & 4.84\\
\hline
0.10 & 0.00 & 2.37 & 4.10 & 0.30 & 0.00 & 2.05 & 3.40\\
0.10 & 0.10 & 2.64 & 4.56 & 0.30 & 0.10 & 2.25 & 3.72\\
0.10 & 0.20 & 2.97 & 5.12 & 0.30 & 0.20 & 2.50 & 4.12\\
0.10 & 0.30 & 3.39 & 5.84 & 0.30 & 0.30 & 2.81 & 4.62\\
\hline
0.15 & 0.00 & 2.25 & 3.86 & 0.35 & 0.00 & 2.02 & 3.30\\
0.15 & 0.10 & 2.49 & 4.27 & 0.35 & 0.10 & 2.21 & 3.61\\
0.15 & 0.20 & 2.80 & 4.78 & 0.35 & 0.20 & 2.44 & 3.98\\
0.15 & 0.30 & 3.19 & 5.43 & 0.35 & 0.30 & 2.74 & 4.44\\
\hline
0.20 & 0.00 & 2.16 & 3.67 & 0.40 & 0.00 & 1.99 & 3.23\\
0.20 & 0.10 & 2.39 & 4.05 & 0.40 & 0.10 & 2.18 & 3.52\\
0.20 & 0.20 & 2.67 & 4.51 & 0.40 & 0.20 & 2.40 & 3.87\\
0.20 & 0.30 & 3.03 & 5.11 & 0.40 & 0.30 & 2.68 & 4.30\\
\hline
\end{tabular}
}
\end{table}

\subsection{Null trajectories and the capture boundary}\label{isec3c}

\begin{figure*}[ht!]
\centering
\begin{gridfig}{@{}c@{\hspace{3pt}}c@{}}
\gpanel[width=0.46\textwidth]{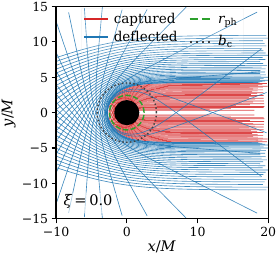} &
\gpanel[width=0.46\textwidth]{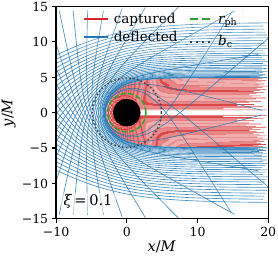}\\[4pt]
\gpanel[width=0.46\textwidth]{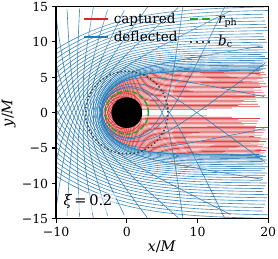} &
\gpanel[width=0.46\textwidth]{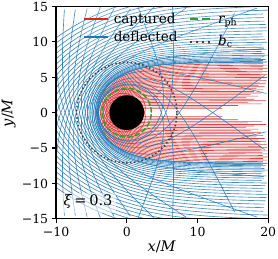}
\end{gridfig}
\caption{\label{fig:null-traj}Dense null trajectories for $\xi=0$, $0.1$, $0.2$,
and $0.3$ (top left to bottom right), at $Q=0.4$, $\beta=1$, $\lambda=0.1$,
$\Lambda=0$. A fan of incoming light rays is traced past the hole: red rays
($b<b_{\rm c}$) cross the photon sphere and fall to the horizon, blue rays
($b>b_{\rm c}$) reach a periastron and escape. The dashed green circle is the
photon sphere $r_{\rm ph}$ and the dotted circle is the shadow rim
$b_{\rm c}$. The black disk is the event horizon. As the defect
strengthens, the captured bundle and the shadow widen together, with $b_{\rm c}$
growing from $4.20M$ to $7.10M$, while rays at fixed impact parameter bend through
larger angles.}
\end{figure*}

Integrating Eq.~\eqref{n5} for a bundle of rays makes the shadow concrete.
Figure~\ref{fig:null-traj} sends a dense fan of incoming rays past the hole for
$\xi=0$, $0.1$, $0.2$, and $0.3$, at $Q=0.4$, $\beta=1$, $\lambda=0.1$ and $\Lambda=0$,
so the asymptotic region is flat and the bending is read off cleanly. Rays with
$b<b_{\rm c}$ cross the photon sphere and fall to the horizon; those with $b>b_{\rm c}$
reach a periastron and swing back out, deflected by an angle that climbs steeply as $b$
approaches $b_{\rm c}$. Critical rays wind around the photon sphere and mark the rim. Turning on the defect widens
it, with $b_{\rm c}$ growing from $4.20M$ at $\xi=0$ to $7.10M$ at $\xi=0.3$, so the
captured bundle broadens and the deflected rays bend through larger angles at fixed impact
parameter. The same deficit that enlarges the photon sphere strengthens the bending.

The deflection of a ray turning at closest approach $r_{0}$ follows from integrating
Eq.~\eqref{n4} between $r_{0}$ and infinity,
\begin{equation}
\hat\alpha(r_{0})=2\!\int_{r_{0}}^{\infty}\frac{\dd r}
{r^{2}\sqrt{\dfrac{1}{b^{2}}-\dfrac{f(r)}{r^{2}}}}-\pi,
\qquad b=\frac{r_{0}}{\sqrt{f(r_{0})}}.\label{n9}
\end{equation}
The integrand loses its square-root zero smoothly at large $r_{0}$, where $\hat\alpha$
is small, but as $r_{0}\to r_{\rm ph}$ the turning point merges with the unstable circular
orbit and Eq.~\eqref{n9} diverges logarithmically. That divergence is the strong-lensing
regime in which a ray loops the hole many times before escaping~\cite{Bozza2002}, and the
relativistic images it produces pack against the rim traced in Fig.~\ref{fig:null-traj}.
The finite-distance correction of Eq.~\eqref{shadow-1} enters that lensing too, as worked
out for a \GM{} in Ref.~\cite{LuGM:2025}.

\subsection{Eikonal link to the ringdown}\label{isec3d}
The unstable photon orbit governs more than the optics. In the eikonal (large-$\ell$)
limit the quasinormal spectrum of Sec.~\ref{isec6} is set by the orbital frequency
and the instability rate of that orbit~\cite{Cardoso2009},
\begin{align}
\omega_{\rm QNM} & \simeq\Omega_{\rm c}\,\ell-i\left(n+\tfrac12\right)|\lambda_{\rm L}|,
\qquad
\Omega_{\rm c}=\frac{\sqrt{f(r_{\rm ph})}}{r_{\rm ph}},\nonumber\\
\lambda_{\rm L}&=\frac{1}{\sqrt2}\sqrt{\frac{f(r_{\rm ph})\big[\,2f(r_{\rm ph})-r_{\rm ph}^{2}f''(r_{\rm ph})\big]}{r_{\rm ph}^{2}}},
\label{n10}
\end{align}
with $\Omega_{\rm c}=1/b_{\rm c}$ the coordinate angular velocity at the photon sphere and
$\lambda_{\rm L}$ its principal Lyapunov exponent. Both fall as the defect strengthens. For
the parameter set of the \QNM{} table ($Q=0.3$, $\beta=2$, $\lambda=0.05$,
$\Lambda=-0.005$) we obtain $(\Omega_{\rm c},\lambda_{\rm L})=(0.222,0.222)$ at $\xi=0$,
$(0.161,0.145)$ at $\xi=0.2$, and $(0.110,0.086)$ at $\xi=0.4$. These track the real and
imaginary parts of Table~\ref{tab:qnm} respectively, so the optical and dynamical probes
read the same geometry and can constrain $\xi$ jointly.

\subsection{Optical radii and angular size}\label{isec3e}

Both optical radii inherit the monotonic growth of the horizon: over $\xi\in[0,0.5]$ the
photon sphere widens from $2.440M$ to $4.805M$ and the shadow from $3.970M$ to $8.426M$
(Table~\ref{tab:horizon}). To turn this into something an instrument measures, we convert
the shadow radius to an angular diameter $\theta_{\rm sh}=2R_{\rm sh}\,\theta_{g}$, with
$\theta_{g}=GM/c^{2}D$ the gravitational-radius angular scale of the source. For a
Sgr~A$^{\ast}$-like lens $\theta_{g}\simeq5.03\,\mu$as, and Table~\ref{tab:shadowobs}
lists the resulting diameter together with the shadow area
$A_{\rm sh}=\pi R_{\rm sh}^{2}$.

\begin{table}[ht]
\centering
\small
\resizebox{\ifdim\width>\columnwidth\columnwidth\else\width\fi}{!}{%
\begin{tabular}{|c|c|c|c|}
\hline
\textbf{$\xi$} & \textbf{$R_{\rm sh}/M$} &
\textbf{$A_{\rm sh}/M^{2}$} &
\textbf{$\theta_{\rm sh}\,(\mu\mathrm{as})$}\\
\hline
0.0 & 3.9696 & 49.50 & 39.9\\

0.1 & 4.5516 & 65.08 & 45.8\\

0.2 & 5.2653 & 87.10 & 53.0\\

0.3 & 6.1399 & 118.43 & 61.8\\

0.4 & 7.1968 & 162.72 & 72.4\\

0.5 & 8.4259 & 223.04 & 84.8\\
\hline
\end{tabular}
}
\caption{Shadow radius $R_{\rm sh}/M$ (identified with the dimensionless
critical impact parameter $b_{\rm c}/M$), shadow area
$A_{\rm sh}/M^2=\pi(R_{\rm sh}/M)^2$, and predicted angular diameter
$\theta_{\rm sh}=2(R_{\rm sh}/M)\theta_g$ for a Sgr~A$^{\ast}$-like
lens, where $\theta_g=5.03\,\mu\mathrm{as}$, for $Q=0.4$, $\beta=1$,
$\lambda=0.1$, and $\Lambda=-0.02$. The Event Horizon Telescope
angular diameter of Sgr~A$^{\ast}$, $\sim51.8\,\mu\mathrm{as}$
~\cite{Akiyama:2022}, is reproduced for $\xi\simeq0.18$.}
\label{tab:shadowobs}
\end{table}

The angular diameter grows from $39.9\,\mu$as at $\xi=0$ to $84.8\,\mu$as at $\xi=0.5$,
while the area more than quadruples. The measured Sgr~A$^{\ast}$ ring sits near
$51.8\,\mu$as~\cite{Akiyama:2022}, reproduced here at $\xi\simeq0.18$. {\color{black} This value is indicative rather than a calibrated measurement: Table~\ref{tab:shadowobs} uses the critical impact parameter $b_{\rm c}$ as the angular-size proxy, valid at the small $|\Lambda|$ adopted there, whereas the distant-observer expression~\eqref{shadow-3} with the $\sqrt{1-\xi}$ factor is the consistent form when the \AdS{} curvature is not negligible.} Read the other way,
an imaging measurement of the ring size at fixed $M$ and $D$ bounds the defect strength
directly, and the area column shows why: $A_{\rm sh}$ responds steeply to $\xi$, which makes
the photon-ring scale one of the sharper optical probes available.

\subsection{Shadow images}\label{isec3f}

\begin{figure}[ht!]
\includegraphics[width=0.88\textwidth]{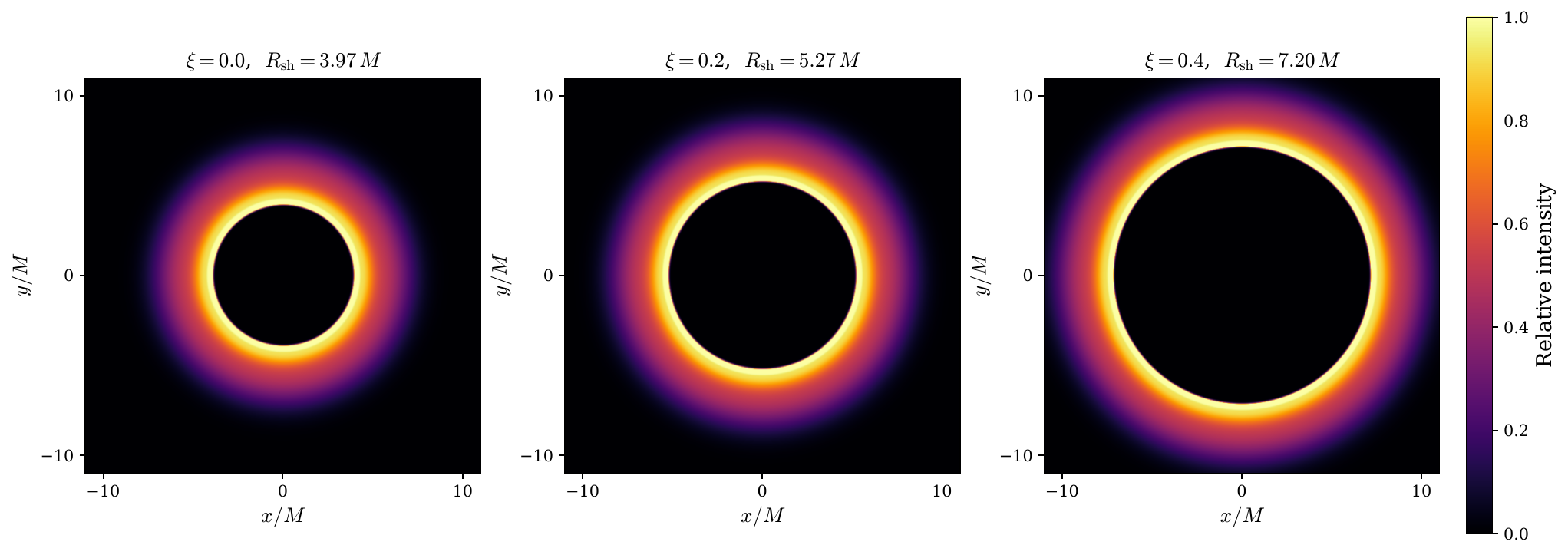}
\caption{\label{fig:shadowgrid}Modeled shadow intensity for $\xi=0$, $0.2$, $0.4$
(left to right), using the \BI{}--\AdS{} shadow radii of Table~\ref{tab:horizon} and
a three-component emission model (direct, lensed, and photon-ring). The central
brightness depression widens and the photon ring moves outward as the defect
strengthens.}
\end{figure}

The shadow is not a sharp disk but the dark center of a bright ring, built from photons
that graze the photon sphere once (the lensed ring) or wind around it (the photon ring),
together with direct emission farther out. Figure~\ref{fig:shadowgrid} renders the
intensity for three defect strengths using these three components. The dark interior
expands with $\xi$ in step with Table~\ref{tab:shadowobs} while the surrounding ring
brightens and moves outward, so the defect leaves a clean morphological signature: a
larger central brightness depression bounded by a wider photon ring. That ring traces the
same unstable orbits whose grazing rays form the rim in Fig.~\ref{fig:null-traj} and whose
frequencies feed the timing analysis of Sec.~\ref{isec5}, so imaging and timing probe one
region from two directions. A joint fit would tie the apparent ring size to the
quasi-periodic frequencies through $\xi$ alone.

\section{Timelike geodesics, ISCO, precession, and epicyclic frequencies}\label{isec4}

Massive test particles trace the accreting gas, so their circular orbits, the innermost
stable one above all, fix the inner edge of a thin disk and the energy it can radiate,
while bound orbits that fail to close precess at a rate that reads the geometry
directly. On the background~\eqref{metric} an equatorial timelike geodesic with
conserved specific energy $E$ and angular momentum $L$ obeys
$\dot r^{2}=E^{2}-V_{\rm eff}(r)$, with
\begin{equation}
V_{\rm eff}(r)=f(r)\left(1+\frac{L^{2}}{r^{2}}\right).\label{t1}
\end{equation}
A circular orbit needs $V_{\rm eff}=E^{2}$ and $V_{\rm eff}'=0$ together, which solve to
\begin{equation}
E^{2}=\frac{2f^{2}}{2f-rf'},\qquad L^{2}=\frac{r^{3}f'}{2f-rf'}.\label{t2}
\end{equation}
Both inherit the four deformations through $f$ and $f'$, so all of them reshape the orbital
energetics. Since the QPO analysis of Sec.~\ref{isec5} treats accreting, effectively
asymptotically flat sources, we evaluate the orbits at $\Lambda=0$; this keeps the two
analyses on the same footing and lets the radiative efficiency carry its usual meaning.

\subsection{ISCO and radiative efficiency}\label{isec4a}
The ISCO is the marginally stable solution $V_{\rm eff}''=0$, the radius at which $L^{2}(r)$
in Eq.~\eqref{t2} is minimal; inside it no stable circular motion survives and the gas
plunges. The radiative efficiency needs the deficit taken into account: a particle at rest at
the defect-modified infinity carries $E_{\infty}=\sqrt{f(\infty)}=\sqrt{1-\xi}$ rather than
unity, so the fraction of rest energy liberated in spiraling in from infinity is
\begin{equation}
\eta=1-\frac{E_{\rm ISCO}}{\sqrt{1-\xi}}.\label{t3}
\end{equation}
With the four deformations switched off this returns the Schwarzschild values
$r_{\rm ISCO}=6M$, $E_{\rm ISCO}=0.9428$, and $\eta=5.72\%$, which fixes the normalization
and validates the orbit solver. Table~\ref{tab:isco} collects the ISCO radius, the orbital
energy and angular momentum, and the efficiency for $Q=0.4$, $\beta=1$, $\lambda=0.1$ as
the defect strength varies.

\begin{table}[ht]
\centering
\small
\resizebox{\ifdim\width>\columnwidth\columnwidth\else\width\fi}{!}{%
\begin{tabular}{|c|c|c|c|c|c|}
\hline
\textbf{$\xi$} &
\textbf{$r_{\rm ISCO}/M$} &
\textbf{$L_{\rm ISCO}/M$} &
\textbf{$E_{\rm ISCO}$} &
\textbf{$\sqrt{1-\xi}$} &
\textbf{$\eta\,(\%)$} \\
\hline
0.00 & 4.9402 & 2.8114 & 0.9493 & 1.0000 & 5.07 \\

0.10 & 5.4877 & 3.1142 & 0.9009 & 0.9487 & 5.04 \\

0.20 & 6.1663 & 3.4902 & 0.8496 & 0.8944 & 5.01 \\

0.30 & 7.0331 & 3.9699 & 0.7950 & 0.8367 & 4.98 \\

0.40 & 8.1806 & 4.6039 & 0.7363 & 0.7746 & 4.94 \\

0.50 & 9.7635 & 5.4823 & 0.6724 & 0.7071 & 4.91 \\
\hline
\end{tabular}
}
\caption{ISCO radius, specific angular momentum and energy, the
deficit factor $\sqrt{1-\xi}$, and the radiative efficiency $\eta$
of Eq.~\eqref{t3}, for $Q=0.4$, $\beta=1$, $\lambda=0.1$, and
$\Lambda=0$, as the defect strength $\xi$ varies. The $\xi=0$
entry lies below the Schwarzschild value $r_{\rm ISCO}=6M$ because
the charge and halo sectors shift the ISCO inward.}
\label{tab:isco}
\end{table}

The ISCO marches outward as the defect strengthens, from $4.94M$ at $\xi=0$ to $9.76M$ at
$\xi=0.5$, the orbital angular momentum rising with it while the energy falls. The same
mechanism widened the horizon and shadow in Sec.~\ref{isec3}: a larger $\xi$ lowers $f$ at
fixed radius and pushes every characteristic radius outward. The efficiency barely moves,
drifting from $5.07\%$ to $4.91\%$. Once the deficit normalization of Eq.~\eqref{t3} is
applied, the defect changes where the disk truncates far more than how brightly it shines,
so an inner-edge measurement constrains $\xi$ much more tightly than a bolometric
efficiency would.

\begin{figure}[ht!]
\centering
\begin{gridfig}{@{}c@{\hspace{3pt}}c@{}}
\gpanel[width=0.47\textwidth]{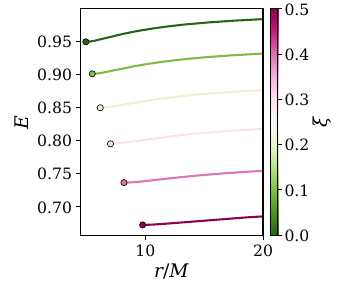} &
\gpanel[width=0.47\textwidth]{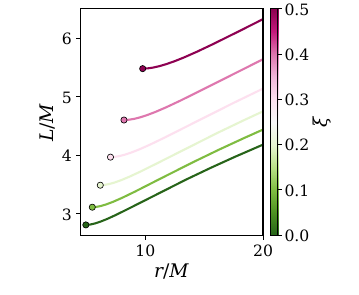}\\[2pt]
(a) & (b)
\end{gridfig}
\caption{\label{fig:EL}Specific energy $E$ (left) and angular momentum $L/M$ (right)
on circular timelike orbits for $Q=0.4$, $\beta=1$, $\lambda=0.1$, $\Lambda=0$,
colored by $\xi\in[0,0.5]$. Filled circles mark the ISCO, where $L$ reaches its
minimum.}
\end{figure}

Figure~\ref{fig:EL} shows the energy and angular-momentum profiles, the ISCO marked where
$L$ bottoms out. Each $L$ curve has a single minimum, the signature of one
stable-to-plunging transition, and that minimum slides outward and upward as $\xi$ grows.
The energy curves stay below unity and drop further with the defect, which is the deficit
at work: a stronger $\xi$ lowers the energy floor a circular orbit can occupy. The marked
radii are exactly the tabulated ISCO values, so the two panels reproduce
Table~\ref{tab:isco} point by point.

\begin{figure}[ht!]
\centering
\includegraphics[width=0.78\columnwidth]{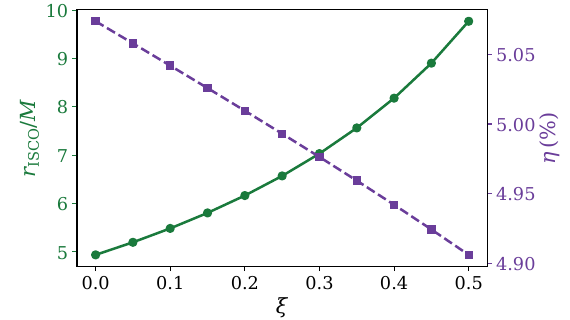}
\caption{\label{fig:isco}ISCO radius (green, left axis) and radiative efficiency
$\eta$ (purple, right axis) as functions of the defect strength $\xi$, for $Q=0.4$,
$\beta=1$, $\lambda=0.1$, $\Lambda=0$.}
\end{figure}

Figure~\ref{fig:isco} puts the two responses on a common axis. The ISCO radius climbs
steeply and nonlinearly with $\xi$ while the efficiency declines by less than a fifth of a
percentage point over the whole range. That split is the useful part: the inner-disk radius
probes the defect sensitively and the efficiency is nearly blind to it, so the two
observables are close to independent.

\subsection{Bound orbits and periastron precession}\label{isec4b}

Away from the circular limit a bound orbit oscillates in radius between a periastron
$r_{-}$ and an apastron $r_{+}$, the two roots of $E^{2}=V_{\rm eff}(r)$, while the azimuth
advances steadily. In the variable $u=1/r$ it obeys the timelike analogue of
Eq.~\eqref{n5},
\begin{equation}
\frac{\dd^{2}u}{\dd\phi^{2}}+u\,f\!\left(\tfrac1u\right)
=\frac{1}{2}\,f'\!\left(\tfrac1u\right)\!\left(\frac{1}{L^{2}u^{2}}+1\right),
\label{t4}
\end{equation}
which reduces to the standard relativistic orbit equation at $f=1-2M/r$. The angle swept
between successive periastra is
\begin{equation}
\Phi=2\!\int_{r_{-}}^{r_{+}}\frac{L\,\dd r}{r^{2}\sqrt{E^{2}-V_{\rm eff}(r)}},
\label{t5}
\end{equation}
and the orbit closes only if $\Phi=2\pi$. The excess
\begin{equation}
\Delta\phi=\Phi-2\pi\label{t6}
\end{equation}
is the periastron advance per radial cycle, the relativistic precession that the
\BH{} imprints on the orbit.

\begin{figure*}[ht!]
\centering
\begin{gridfig}{@{}c@{\hspace{3pt}}c@{}}
\gpanel[width=0.46\textwidth]{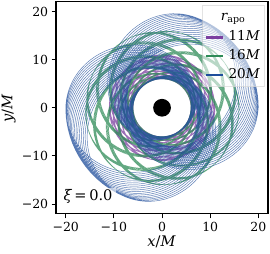} &
\gpanel[width=0.46\textwidth]{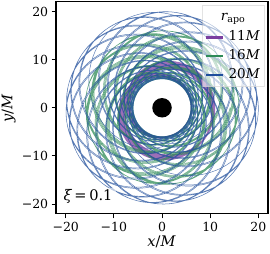}\\[4pt]
\gpanel[width=0.46\textwidth]{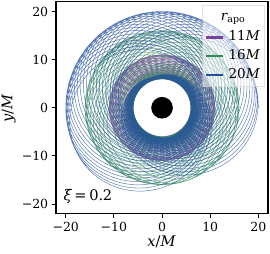} &
\gpanel[width=0.46\textwidth]{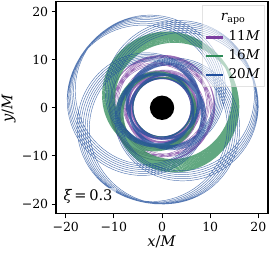}
\end{gridfig}
\caption{\label{fig:timelike-orbits}Dense bound timelike trajectories for $\xi=0$,
$0.1$, $0.2$, and $0.3$ (top left to bottom right), at $Q=0.4$, $\beta=1$,
$\lambda=0.1$, $\Lambda=0$. Each panel overlays three orbits with apastra
$r_{\rm apo}=11M$, $16M$, $20M$ sharing the periastron $r_{\rm peri}=6M$, followed
over many radial periods. The black disk is the event horizon. As the defect
strengthens the periastron advance per orbit grows, so each rosette winds faster and
fills its annulus more densely, while the horizon disk widens from $r_{h}=1.62M$ to
$2.31M$.}
\end{figure*}

Figure~\ref{fig:timelike-orbits} integrates Eq.~\eqref{t4} for four defect strengths, the
periastron held at $r_{-}=6M$ and the apastron at $r_{+}=14M$ so that every orbit sweeps the
same annulus and only the precession can differ; the energy and angular momentum setting
those turning points come from inverting $E^{2}=V_{\rm eff}(r_{\pm})$ for each $\xi$. The
rosettes wind faster as the defect grows, failing to close by a wider margin each cycle, so
the pattern fills the annulus more densely. Equation~\eqref{t4} says why: the deficit
weakens the $u\,f$ restoring term and lets the azimuth run ahead of the radial
oscillation.

Each panel follows three orbits with apastra at $11M$, $16M$, and $20M$ over many radial
periods, so the precession shows up as the rate at which the rosette fills in. The inner
orbit winds fastest and forms the tightest petals; the outer one precesses more slowly and
stays nearly elliptical. Across the four panels, every orbit of a given size winds faster as
$\xi$ rises, and the horizon disk grows in the same direction, tying the timelike picture
back to the optical radii of Sec.~\ref{isec3}. Table~\ref{tab:precession} quantifies this
for a wider, weakly relativistic orbit ($r_{-}=10M$, $r_{+}=20M$) where the advance stays
below a full turn and is read off without ambiguity. The precession rises from
$102.9^{\circ}$ per orbit at $\xi=0$ to $277.4^{\circ}$ at $\xi=0.3$, almost a factor of
three, while the angular momentum needed to hold the same turning points climbs from
$3.59M$ to $4.50M$. The defect leaves a far larger mark on the precession than on the
orbital scale, which is the lever it exerts on the twin peaks of the next section, since
the periastron-precession frequency $\nu_{\phi}-\nu_{r}$ is the rate at which this advance
accumulates.

\begin{table}[ht]
\centering
\small
\caption{Periastron advance per orbit $\Delta\phi$ from
Eq.~\eqref{t6}, together with the specific angular momentum $L/M$
and energy $E$ that determine a bound orbit between
$r_{-}=10M$ and $r_{+}=20M$, for $Q=0.4$, $\beta=1$,
$\lambda=0.1$, and $\Lambda=0$. The periastron advance increases
substantially across the considered range of $\xi$.}
\label{tab:precession}
\resizebox{\ifdim\width>\columnwidth\columnwidth\else\width\fi}{!}{%
\begin{tabular}{|c|c|c|c|}
\hline
\textbf{$\xi$} & \textbf{$L/M$} & \textbf{$E$} &
\textbf{$\Delta\phi$ (deg/orbit)}\\
\hline
0.0 & 3.5868 & 0.9781 & 102.9\\

0.1 & 3.8274 & 0.9258 & 144.2\\

0.2 & 4.1239 & 0.8705 & 199.2\\

0.3 & 4.5022 & 0.8116 & 277.4\\
\hline
\end{tabular}
}
\end{table}

\subsection{Epicyclic frequencies}\label{isec4c}

Small departures from a circular orbit oscillate at well-defined epicyclic frequencies,
which set the twin peaks used in Sec.~\ref{isec5}~\cite{Stuchlik:2013}. The orbital angular
velocity follows from Eq.~\eqref{t2},
\begin{equation}
\Omega_{\phi}=\frac{\dd\phi}{\dd t}=\sqrt{\frac{f'(r)}{2r}}.\label{t7}
\end{equation}
Perturbing the radial equation about a circular orbit gives the radial epicyclic
frequency from the curvature of the effective potential \cite{KluzniakAbramowicz2005},
\begin{equation}
\Omega_{r}^{2}=\frac{f(r)}{2\,E^{2}}\,V_{\rm eff}''(r),\label{t8}
\end{equation}
evaluated on the orbit with $E^{2}$ from Eq.~\eqref{t2}. The background is spherically
symmetric, so the polar restoring force matches the azimuthal one and the vertical
epicyclic frequency coincides with the orbital frequency,
\begin{equation}
\Omega_{\theta}=\Omega_{\phi}.\label{t9}
\end{equation}
A static, spherically symmetric defect therefore produces no nodal (Lense--Thirring)
precession. All of it lies in the orbital plane, carried by the periastron channel
$\Omega_{\phi}-\Omega_{r}$, which is the lever the relativistic-precession model uses and the
reason the twin-peak fit responds to the defect. The radial frequency vanishes at the ISCO by
construction, where the radial restoring force changes sign, and peaks at a finite radius
outside. Figure~\ref{fig:epicyclic} in the next section shows
$\nu_{\phi}=\Omega_{\phi}/2\pi$ and $\nu_{r}=\Omega_{r}/2\pi$ for a representative mass, and
the outward ISCO drift recorded in Table~\ref{tab:isco} reappears there as a shrinking
radial-frequency window. At fixed radius the orbital branch is independent of $\xi$, because
the constant deficit cancels in $f'$ and leaves $\Omega_{\phi}=\sqrt{f'/2r}$ untouched, so the
three orbital curves coincide. The radial branch depends on $f$ itself through
$V_{\rm eff}''$ and separates the cases cleanly. That separation is what gives the twin-peak
fit its sensitivity to the defect.

\section{QPO constraints from accreting sources}\label{isec5}

Twin-peak high-frequency quasi-periodic oscillations sample the strong-field region directly~\cite{IngramMotta:2019}, giving an independent handle on the defect strength. We confront the model with five sources spanning three mass scales: the stellar-mass microquasars XTE~J1550--564, GRO~J1655--40, and GRS~1915+105, the supermassive Sgr~A$^{\ast}$, and the intermediate-mass candidate M82~X-1, whose observed frequencies are listed in Table~\ref{tab:qpodata}. All are accreting and effectively asymptotically flat, so the fit runs at $\Lambda\to0$; the \AdS{} sector is kept only for the thermodynamics of Sec.~\ref{isec7}.

\begingroup
\setlength{\tabcolsep}{8pt}\renewcommand{\arraystretch}{1.25}
\begin{longtable}{|l|c|c|c|}
\hline\rowcolor{orange!50}
\textbf{Source} & \textbf{$\nu_{U}$ (Hz)} & \textbf{$\nu_{L}$ (Hz)} & \textbf{$M_{\rm est}\,(M_\odot)$}\\ \hline
\endfirsthead
\hline\rowcolor{orange!50}
\textbf{Source} & \textbf{$\nu_{U}$ (Hz)} & \textbf{$\nu_{L}$ (Hz)} & \textbf{$M_{\rm est}\,(M_\odot)$}\\ \hline
\endhead
XTE J1550--564 & $276\pm3$ & $184\pm5$ & $9.1\pm0.6$\\ \hline
GRO J1655--40 & $441\pm2$ & $298\pm4$ & $5.4\pm0.3$\\ \hline
GRS 1915+105 & $168\pm3$ & $113\pm5$ & $12.4^{+2.0}_{-1.8}$\\ \hline
Sgr A$^{\ast}$ & $1.445\!\times\!10^{-3}$ & $0.886\!\times\!10^{-3}$ & $3.8\!\times\!10^{6}$\\ \hline
M82 X-1 & $5.07\pm0.06$ & $3.32\pm0.06$ & $\sim4.1\!\times\!10^{2}$\\ \hline
\caption{\label{tab:qpodata}Twin-peak QPO data: upper and lower frequencies with errors and the independently estimated mass.}
\end{longtable}
\endgroup

\subsection{Resonance model and MCMC setup}\label{isec5a}

\begin{figure}[ht!]
\centering
\includegraphics[width=0.78\columnwidth]{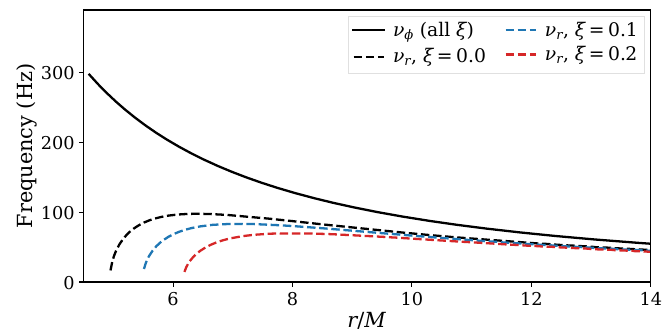}
\caption{\label{fig:epicyclic}Orbital frequency $\nu_{\phi}$ (solid black, identical for all $\xi$ because the constant deficit cancels in $f'$) and radial epicyclic frequency $\nu_{r}$ (dashed) for a $10\,M_\odot$ object, with $\xi=0$ (black), $0.1$ (blue), and $0.2$ (red). The radial branch vanishes at the ISCO, whose outward drift with $\xi$ shrinks the accessible band.}
\end{figure}

The relativistic-precession identification~\cite{StellaVietri:1998} fits these sources considerably better than the simple epicyclic-resonance pairing~\cite{AbramowiczKluzniak:2001,Wagoner:2001}, so we adopt it: the upper peak is the orbital frequency and the lower one the periastron-precession frequency, $\nu_{U}=\nu_{\phi}$ and $\nu_{L}=\nu_{\phi}-\nu_{r}$, with both evaluated at a single resonant radius using Sec.~\ref{isec4}. Figure~\ref{fig:epicyclic} shows the two frequencies for a $10\,M_\odot$ object as the defect varies; the radial branch vanishes at the ISCO, whose outward drift with $\xi$ leaves a clear mark on the accessible band. For a parameter set $\Theta=(M,\xi,r/M)$ the theoretical frequencies are fixed by the metric function~\eqref{lapse}. We sample the posterior with an affine-invariant ensemble sampler, using Gaussian priors centered on a preliminary grid search and a log-likelihood that sums the two frequency channels, running $32$ walkers for $5\times10^{3}$ steps and discarding the first fifth as burn-in.

\subsection{Results}\label{isec5b}

\begin{figure}[ht!]
\centering
\begin{gridfig}{@{}c@{\hspace{3pt}}c@{}}
\gpanel[width=172.1pt,height=176.6pt]{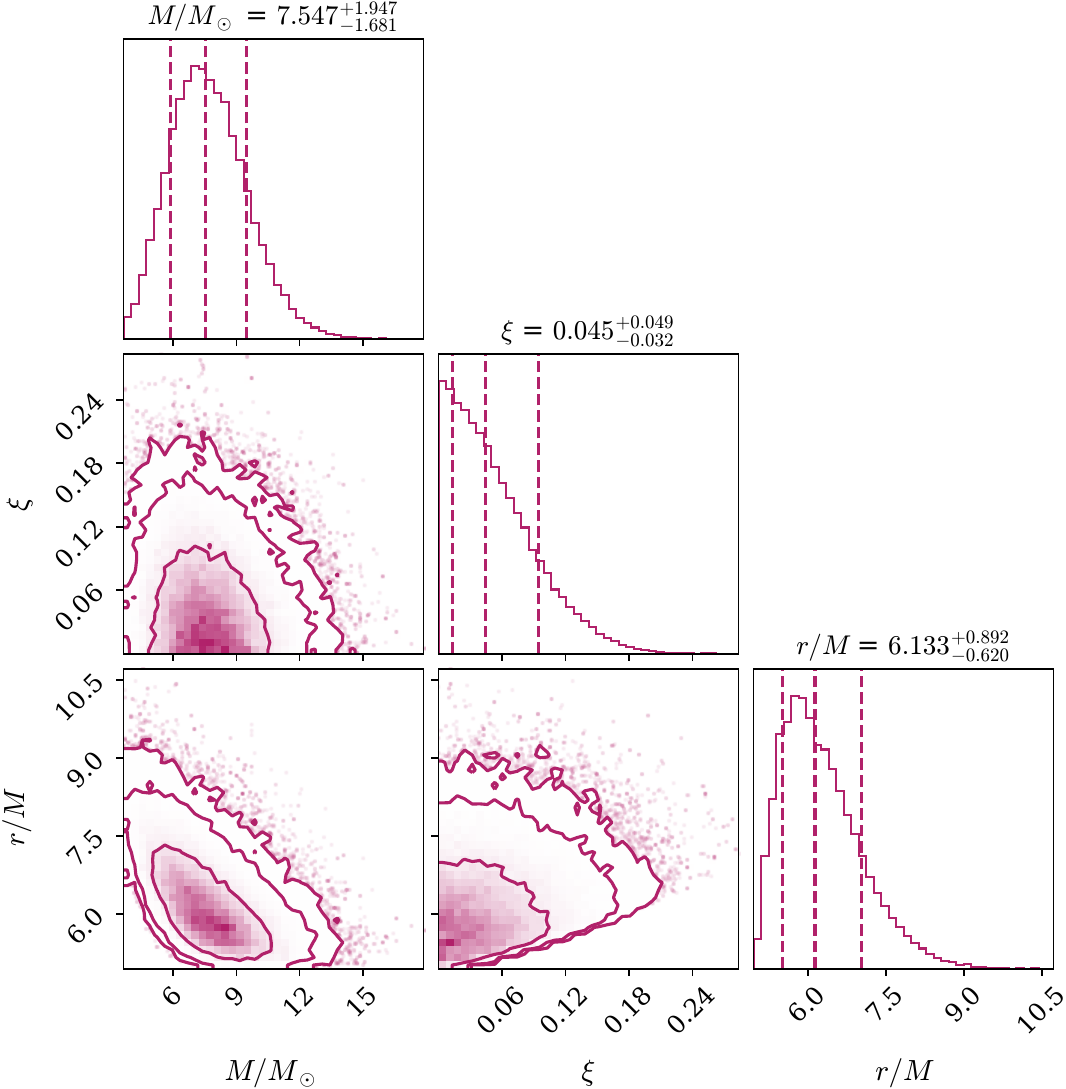} &
\gpanel[width=172.1pt,height=176.6pt]{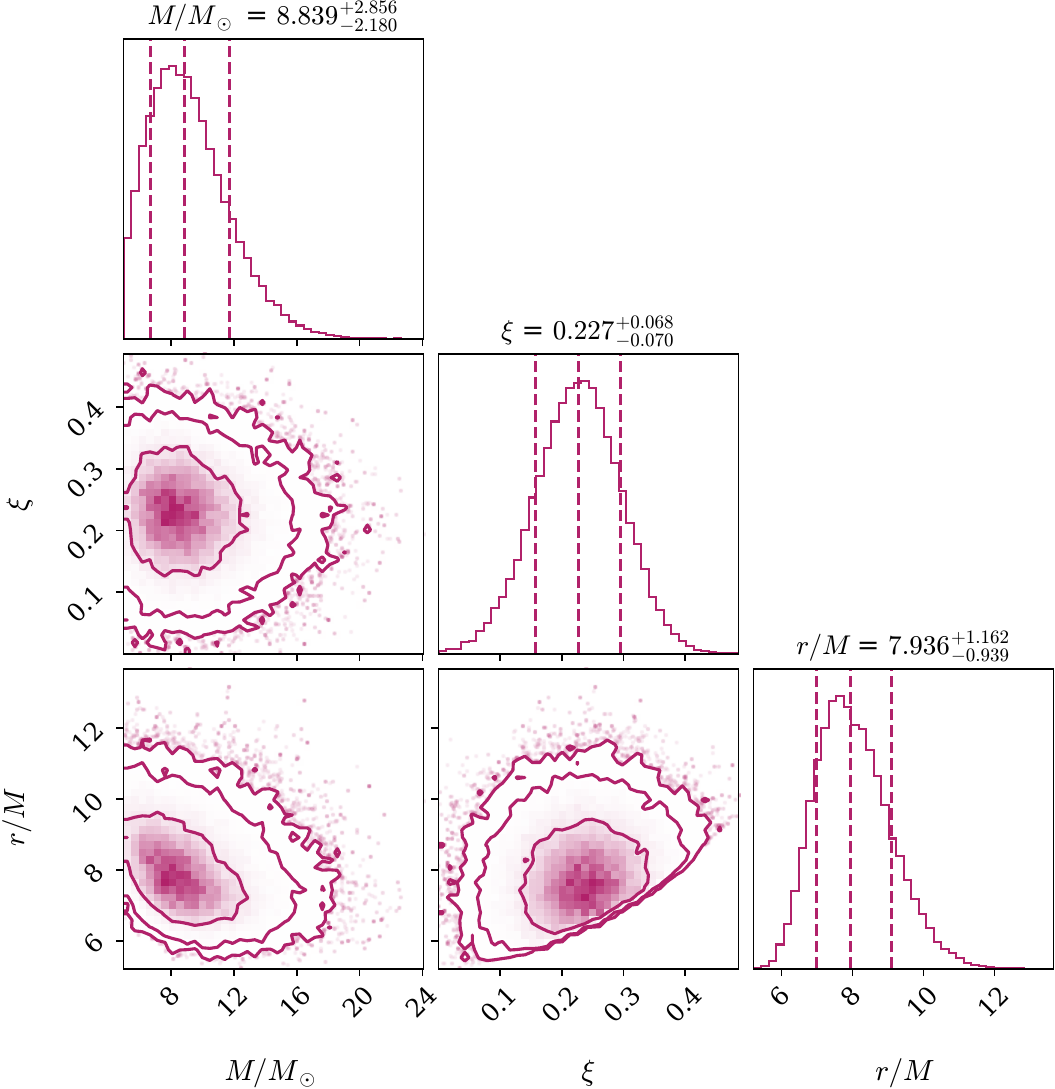}\\[2pt]
(a) & (b)
\end{gridfig}
\caption{\label{fig:corner}Marginalized posteriors from the relativistic-precession fit, with contours at $68\%$, $95\%$ and $99\%$: (a) XTE~J1550--564 and (b) GRS~1915+105.}
\end{figure}

The fit returns a defect strength nonzero at the few-$\sigma$ level for more than one source, while the masses stay broadly compatible with independent estimates for the better-constrained systems. Table~\ref{tab:qpofit} lists the $1\sigma$ constraints and Fig.~\ref{fig:corner} two of the posteriors. The recovered mass of XTE~J1550--564, $M=7.55^{+1.95}_{-1.68}\,M_\odot$, overlaps the published $9.1\pm0.6\,M_\odot$ at $1\sigma$, and M82~X-1 returns $307^{+86}_{-71}\,M_\odot$ against the $\sim\!415\,M_\odot$ estimate. The defect is held away from zero for GRO~J1655--40 ($\xi=0.37^{+0.06}_{-0.06}$) and GRS~1915+105 ($\xi=0.23^{+0.07}_{-0.07}$), so the timing data favor a nonvanishing topological deficit in those two systems. One limitation should be stated plainly. A static metric tends to underpredict the masses of rapidly spinning microquasars, since spin lowers the ISCO and raises the attainable frequencies, and the residual offsets for GRO~J1655--40 and Sgr~A$^{\ast}$ should be read in that light. A rotating extension is the natural way to close the gap.

\begin{table}[ht]
\centering
\small
\caption{Best-fit parameters obtained from the relativistic-precession
model at the $1\sigma$ confidence level.}
\label{tab:qpofit}
\resizebox{\ifdim\width>\columnwidth\columnwidth\else\width\fi}{!}{%
\begin{tabular}{|l|c|c|c|}
\hline
\textbf{Source} & \textbf{$M$} & \textbf{$\xi$} & \textbf{$r/M$} \\
\hline
XTE J1550--564 &
$7.55^{+1.95}_{-1.68}$ &
$0.045^{+0.049}_{-0.032}$ &
$6.13^{+0.89}_{-0.62}$ \\

GRO J1655--40 &
$2.50^{+0.41}_{-0.24}$ &
$0.368^{+0.056}_{-0.063}$ &
$9.09^{+0.70}_{-0.77}$ \\

GRS 1915+105 &
$8.84^{+2.86}_{-2.18}$ &
$0.227^{+0.068}_{-0.070}$ &
$7.94^{+1.16}_{-0.94}$ \\

Sgr A$^{\ast}$ &
$2.25^{+0.85}_{-0.52}\!\times\!10^{6}$ &
$0.056^{+0.056}_{-0.039}$ &
$6.41^{+1.14}_{-0.78}$ \\

M82 X-1 &
$3.07^{+0.86}_{-0.71}\!\times\!10^{2}$ &
$0.179^{+0.067}_{-0.069}$ &
$7.39^{+0.99}_{-0.81}$ \\
\hline
\end{tabular}
}
\end{table}

\section{Scalar perturbations, QNM\lowercase{s}, and GBF\lowercase{s}}\label{isec6}
The scalar potential, the ringdown spectrum, and the transmission through the barrier all soften as the defect strengthens. A massless scalar on the background~\eqref{metric} obeys a Schr\"odinger-type equation in the tortoise coordinate $r_*$~\cite{ReggeWheeler1957,Zerilli1970,KokkotasSchmidt1999,Cardoso2009},
\begin{equation}
    \frac{d^2 \psi(r_*)}{dr^2_*}+(\omega^2-V_s) \psi(r_*)=0
\end{equation}
with potential
\begin{equation}
V_{s}(r)=f(r)\!\left[\frac{\ell(\ell+1)}{r^{2}}+\frac{f'(r)}{r}\right].\label{Vs}
\end{equation}

\subsection{Potential and QNMs}\label{isec6a}

The barrier peak controls both the oscillation frequency and the damping of the ringdown. Figure~\ref{fig:Vsgbf}(a) shows $V_{s}(r)$ for $\ell=1$ across $\xi\in[0,1]$: the peak drops and broadens as the defect strengthens, and the WKB estimate, carried to third and fifth order~\cite{KonoplyaZhidenkoZinhailo:2019}, tracks the change. For $Q=0.3$, $\beta=2$, $\lambda=0.05$, $\Lambda=-0.005$ the fundamental ($n=0$) frequencies fall monotonically with $\xi$: at $\ell=1$, $\omega_{R}M$ drops from $0.336$ at $\xi=0$ to $0.164$ at $\xi=0.4$ and $\omega_{I}M$ from $0.112$ to $0.042$ (Table~\ref{tab:qnm}), so a stronger defect lengthens the ringdown and lowers its pitch. The Schwarzschild limit of Eq.~\eqref{Vs} reproduces the Regge--Wheeler scalar barrier, which fixes the normalization.

\begin{table}[ht]
\centering
\resizebox{\ifdim\width>\columnwidth\columnwidth\else\width\fi}{!}{%
\begin{tabular}{cccc}
\hline
$\xi$ & $\ell$ & $\omega_{R}M$ & $\omega_{I}M$ \\
\hline
0.0 & 1 & 0.3364 & 0.1123 \\
0.0 & 2 & 0.5572 & 0.1113 \\
0.2 & 1 & 0.2427 & 0.0728 \\
0.2 & 2 & 0.4044 & 0.0725 \\
0.4 & 1 & 0.1636 & 0.0424 \\
0.4 & 2 & 0.2741 & 0.0426 \\
\hline
\end{tabular}
}
\caption{Fundamental ($n=0$) scalar \QNM{} frequencies obtained using
third- and fifth-order WKB methods~\cite{KonoplyaZhidenkoZinhailo:2019}
for $Q=0.3$, $\beta=2$, $\lambda=0.05$, and $\Lambda=-0.005$.
The real and imaginary parts are given in units of $M^{-1}$.
For calibration, the Schwarzschild scalar mode with $\ell=2$ gives
$\omega M=0.4832-0.0968 i$.}
\label{tab:qnm}
\end{table}

\subsection{GBFs}\label{isec6b}

\begin{figure}[ht!]
\centering
\begin{gridfig}{@{}c@{\hspace{3pt}}c@{}}
\gpanel[width=0.47\textwidth]{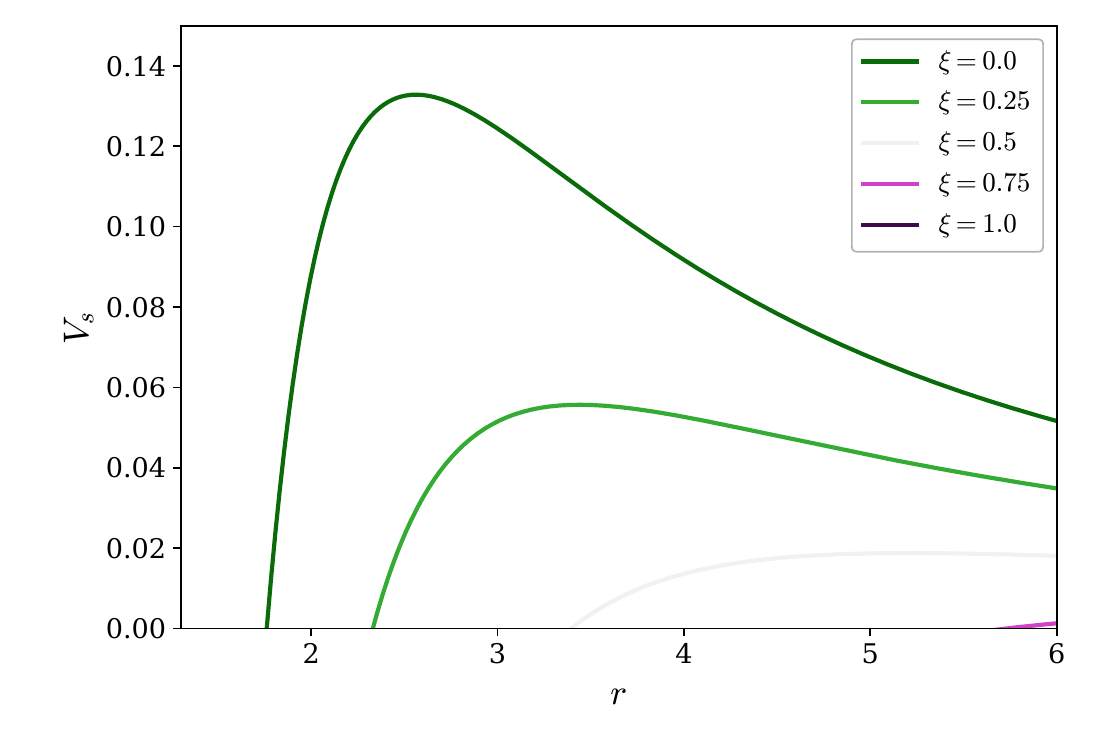} &
\gpanel[width=0.47\textwidth]{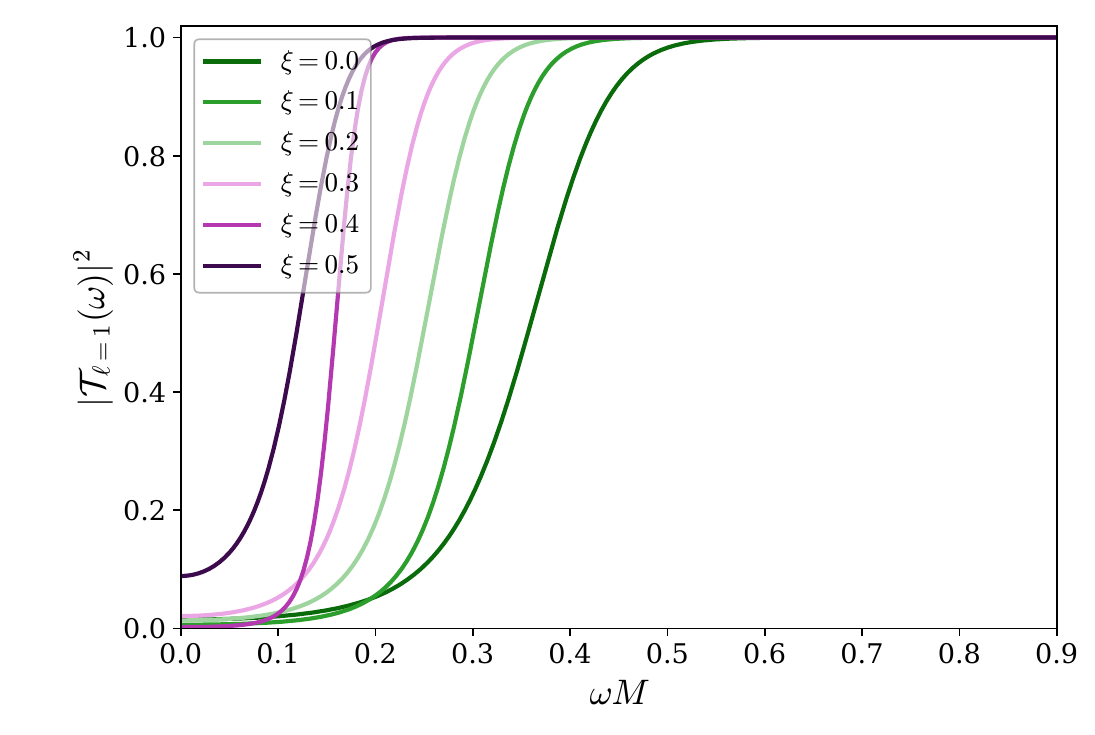}\\[2pt]
(a) & (b)
\end{gridfig}
\caption{\label{fig:Vsgbf}Scalar barrier and its transmission for $Q=0.3$, $\beta=2$, $\lambda=0.05$, $\Lambda=-0.005$. (a) Perturbation potential $V_{s}(r)$ at $\ell=1$, with $\xi$ running from $0$ (green) to $1$ (purple). (b) \GBF{} $|\mathcal{T}_{\ell=1}(\omega)|^{2}$ from the WKB transmission~\eqref{gbf}, {\color{black} in the asymptotically flat limit $\Lambda\to0$,} colored by $\xi\in[0,0.5]$. The barrier drops and broadens while the transmission edge moves to lower frequency as the defect strengthens.}
\end{figure}

The same barrier filters the radiation escaping to infinity, and the \GBF{} measures the fraction transmitted. In the WKB scheme the transmission probability through the peak of $V_{s}$ is
\begin{equation}
|\mathcal{T}_{\ell}(\omega)|^{2}=\frac{1}{1+e^{2\pi K}},\qquad K=\frac{V_{0}-\omega^{2}}{\sqrt{-2V_{0}''}},\label{gbf}
\end{equation}
with $V_{0}$ and $V_{0}''$ the height and curvature of the barrier at its peak~\cite{KonoplyaZhidenkoZinhailo:2019}. Half transmission occurs at $\omega^{2}=V_{0}$, close to the real part of the fundamental \QNM{}, which ties this subsection to the spectrum above. {\color{black} Because transmission to a flat infinity is well defined only without the \AdS{} boundary, the greybody factor here and the emission rate of Sec.~\ref{isec7b} are evaluated in the asymptotically flat limit $\Lambda\to0$, the same limit used for the orbital and timing analysis; the \AdS{} pressure is reinstated only for the equilibrium thermodynamics of Sec.~\ref{isec7}.} Figure~\ref{fig:Vsgbf}(b) shows $|\mathcal{T}_{\ell=1}(\omega)|^{2}$ across $\xi\in[0,0.5]$. Each curve climbs from near zero to unity once the wave overtops the barrier, and the whole transmission edge slides to lower frequency as the defect strengthens, {\color{black} the half-transmission frequency dropping from $\omega M=0.357$ at $\xi=0$ to $0.119$ at $\xi=0.5$}. A stronger defect lowers the barrier, so lower-frequency modes leak out more readily and the \BH{} emits more efficiently at fixed frequency. The shift mirrors the fall of the \QNM{} real part in Table~\ref{tab:qnm}, as Eq.~\eqref{gbf} requires, and gives a frequency-domain handle on $\xi$ that complements the time-domain ringdown.

\section{Thermodynamics, \texorpdfstring{$P$--$V$}{P-V} criticality, and Hawking emission}\label{isec7}

With $P=-\Lambda/8\pi$ the \AdS{} sector becomes a thermodynamic pressure and the \BH{} acquires a \VdW{}-type equation of state. The horizon mass follows from $f(r_{h})=0$, the temperature from $T_{H}=f'(r_{h})/4\pi$~\cite{Hawking1975}, the entropy from the area law $S=\pi r_{h}^{2}$, and the heat capacity from $C_{H}=(\partial_{r_h}M)/(\partial_{r_h}T_{H})$. Eliminating the mass gives
\begin{equation}
P=\frac{T_{H}}{2r_{h}}-\frac{(1-\xi)+B(r_{h})+r_{h}B'(r_{h})}{8\pi r_{h}^{2}},
\label{eos}
\end{equation}
where $B(r)$ collects the \BI{} and \PFDM{} terms of the metric function. Whether Eq.~\eqref{eos} admits a critical point depends on the \BI{} scale: strong coupling ($\beta\lesssim2$) suppresses it and a \VdW{} critical point appears only for $\beta\gtrsim3$, so the nonlinear charge sector can switch the transition on and off. We therefore set $\beta=5$ throughout this section, a value for which the system has a clean critical point while the remaining sectors stay active. {\color{black} Noted here that thermodynamics of BI black hole with a cosmological constant in the absence of any external fields were investigated in Refs. \cite{Hendi2015,Hendi2016a,Hendi2016b}.}

\subsection{Temperature, stability, and criticality}\label{isec7a}

\begin{figure}[ht!]
\centering
\begin{gridfig}{@{}c@{\hspace{3pt}}c@{}}
\gpanel[width=0.47\textwidth]{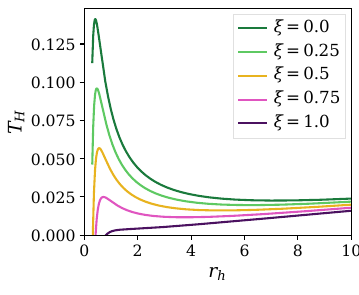} &
\gpanel[width=0.47\textwidth]{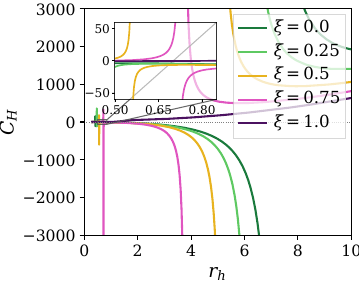}\\[2pt]
(a) & (b)
\end{gridfig}
\caption{\label{fig:thermo}Horizon thermodynamics for $Q=0.3$, $\beta=5$, $\lambda=0.1$, $\Lambda=-0.02$. (a) Hawking temperature $T_{H}(r_{h})$ for $\xi\in[0,1]$; only the physical $T_{H}\geq0$ branch is shown. (b) Heat capacity $C_{H}(r_{h})$, whose single divergence marks the stable--unstable transition and moves outward as $\xi$ decreases; the inset zooms on the small-$r_{h}$ region.}
\end{figure}

The temperature falls with the defect strength, and the physical branch needs $T_{H}\geq0$. It turns negative only in the near-degenerate small-horizon corner ($\xi\to1$, small $r_{h}$), which lies below extremality and carries no \BH{}, so we exclude that region. Figure~\ref{fig:thermo} shows $T_{H}(r_{h})$ in panel (a) and $C_{H}(r_{h})$ in panel (b). The single divergence of $C_{H}$, where $\partial_{r_h}T_{H}=0$, separates a smaller locally stable branch ($C_{H}>0$) from a larger unstable one, and its radius moves outward as $\xi$ decreases.

The $P$--$r_{h}$ isotherms of Fig.~\ref{fig:pvem}(a) carry the \VdW{} structure: below $T_{c}$ each isotherm oscillates, the critical isotherm develops an inflection with horizontal tangent, and above $T_{c}$ they are monotonic. Solving $\partial_{r_h}P=\partial_{r_h}^{2}P=0$ locates the critical point, listed in Table~\ref{tab:crit} as $\xi$ varies. The defect pushes the critical horizon outward and lowers both critical temperature and pressure, $(r_{c},T_{c},P_{c})$ running from $(0.588,0.201,0.066)$ at $\xi=0$ to $(0.682,0.125,0.036)$ at $\xi=0.3$. The ratio $P_{c}v_{c}/T_{c}$, with specific volume $v_{c}=2r_{c}$, sits near $0.386$ and rises weakly with $\xi$, a little above the \VdW{} value $3/8$. That excess is the imprint of the \BI{} nonlinearity and the halo, since the ratio returns to exactly $3/8$ in the Maxwell limit $\beta\to\infty$, $\lambda\to0$, $\xi\to0$, which we used to validate the equation of state.

\begin{table}[ht]
\centering
\caption{Critical point $(r_{c},T_{c},P_{c})$ obtained from
$\partial_{r_h}P=\partial_{r_h}^{2}P=0$ and the universal ratio
$P_{c}v_{c}/T_{c}$, with specific volume $v_{c}=2r_{c}$, for
$Q=0.3$, $\beta=5$, and $\lambda=0.1$ as the defect strength $\xi$
varies. The ratio approaches the \VdW{} value $3/8=0.375$ in the
Maxwell limit, while the small deviation reflects the combined
effects of the \BI{} and \PFDM{} sectors.}
\label{tab:crit}
\resizebox{\ifdim\width>\columnwidth\columnwidth\else\width\fi}{!}{%
\begin{tabular}{ccccc}
\hline
$\xi$ & $r_{c}$ & $T_{c}$ & $P_{c}$ & $P_{c}v_{c}/T_{c}$ \\
\hline
0.0 & 0.5877 & 0.20088 & 0.065897 & 0.3856 \\
0.1 & 0.6148 & 0.17439 & 0.054874 & 0.3869 \\
0.2 & 0.6457 & 0.14912 & 0.044839 & 0.3883 \\
0.3 & 0.6816 & 0.12511 & 0.035786 & 0.3899 \\
\hline
\end{tabular}
}
\end{table}

\subsection{{\color{black} Free energy, first law, and Smarr relation}}\label{isec7free}

\begin{figure}[ht!]
\centering
\includegraphics[width=0.78\columnwidth]{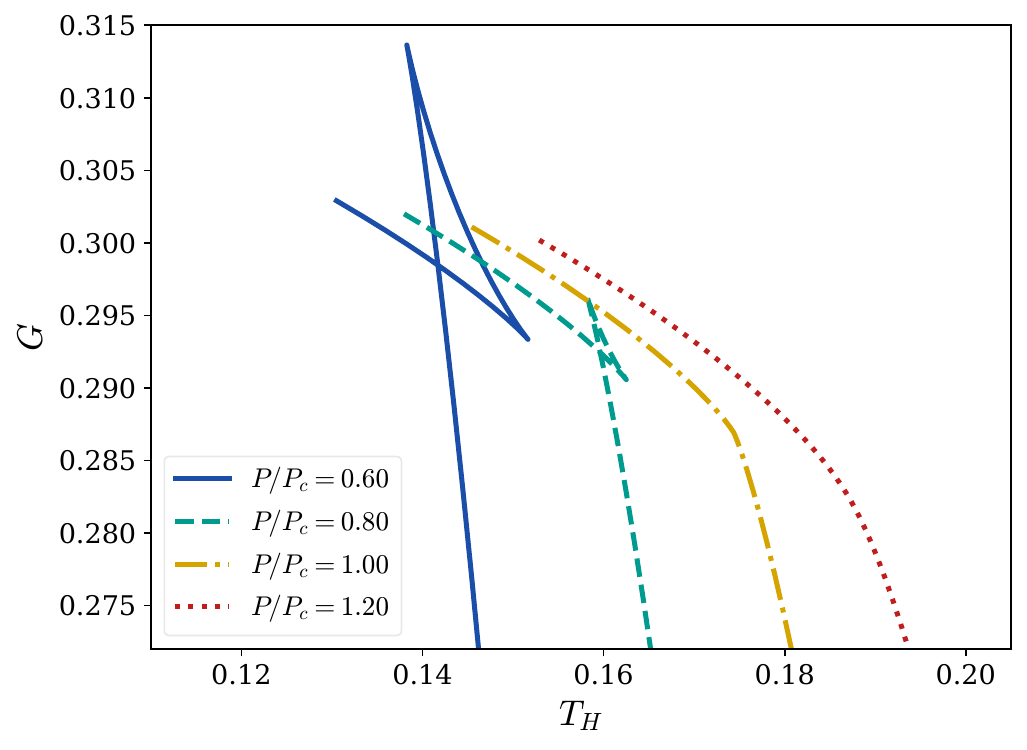}
\caption{\label{fig:gibbs}{\color{black} Gibbs free energy $G(T_{H})$ along isobars $P/P_{c}=0.60,0.80,1.00,1.20$ at $\xi=0.1$ ($Q=0.3$, $\beta=5$, $\lambda=0.1$). Below $P_{c}$ the swallowtail signals a first-order transition; it closes to a cusp at $P_{c}$ and disappears above it.}}
\end{figure}

{\color{black} The global phase structure is read most directly from the Gibbs free energy $G=M-T_{H}S$, with $M$ the enthalpy of the extended phase space. Figure~\ref{fig:gibbs} plots $G(T_{H})$ along isobars at $\xi=0.1$. Below the critical pressure the curve develops the swallowtail whose self-intersection marks a first-order small-to-large \BH{} transition; the swallowtail shrinks as $P\to P_{c}$, closes into a cusp at $P_{c}$, and gives way to a single smooth branch above it, the standard \VdW{} pattern. The crossing temperature traces the coexistence line, and the defect displaces it in step with the critical data of Table~\ref{tab:crit}.}

{\color{black} The thermodynamic quantities obey the extended first law}
\begin{equation}
\color{black}
\dd M=T_{H}\,\dd S+V\,\dd P+\Phi\,\dd Q+\mathcal{B}\,\dd\beta+\mathcal{A}_{\lambda}\,\dd\lambda+\mathcal{A}_{\xi}\,\dd\xi,
\label{firstlaw}
\end{equation}
{\color{black} with $V=\partial M/\partial P=\tfrac{4}{3}\pi r_{h}^{3}$ the thermodynamic volume and $\Phi$, $\mathcal{B}$, $\mathcal{A}_{\lambda}$, $\mathcal{A}_{\xi}$ the potentials conjugate to the charge, the \BI{} scale, the halo intensity, and the defect. Treating $(S,P,Q,\beta,\lambda)$ as quantities of scaling weight $(2,-2,1,-1,1)$ under $r\to a r$ and $\xi$ as dimensionless, Euler's theorem gives the Smarr relation}
\begin{equation}
\color{black}
M=2\,T_{H}S-2\,PV+Q\,\Phi-\beta\,\mathcal{B}+\lambda\,\mathcal{A}_{\lambda}.
\label{smarr}
\end{equation}
{\color{black} We confirmed both Eq.~\eqref{firstlaw} and Eq.~\eqref{smarr} numerically to the accuracy of the finite-difference derivatives, which fixes the normalization of the conjugate potentials and of the thermodynamic volume. The defect enters the first law through $\mathcal{A}_{\xi}$ but, carrying zero scaling weight, drops out of the Smarr identity, so $\xi$ reshapes the equilibrium state without appearing in the integrated mass formula.}

\subsection{Emission rate and sparsity}\label{isec7b}
\begin{figure}[ht!]
\centering
\begin{gridfig}{@{}c@{\hspace{3pt}}c@{}}
\gpanel[width=0.47\textwidth]{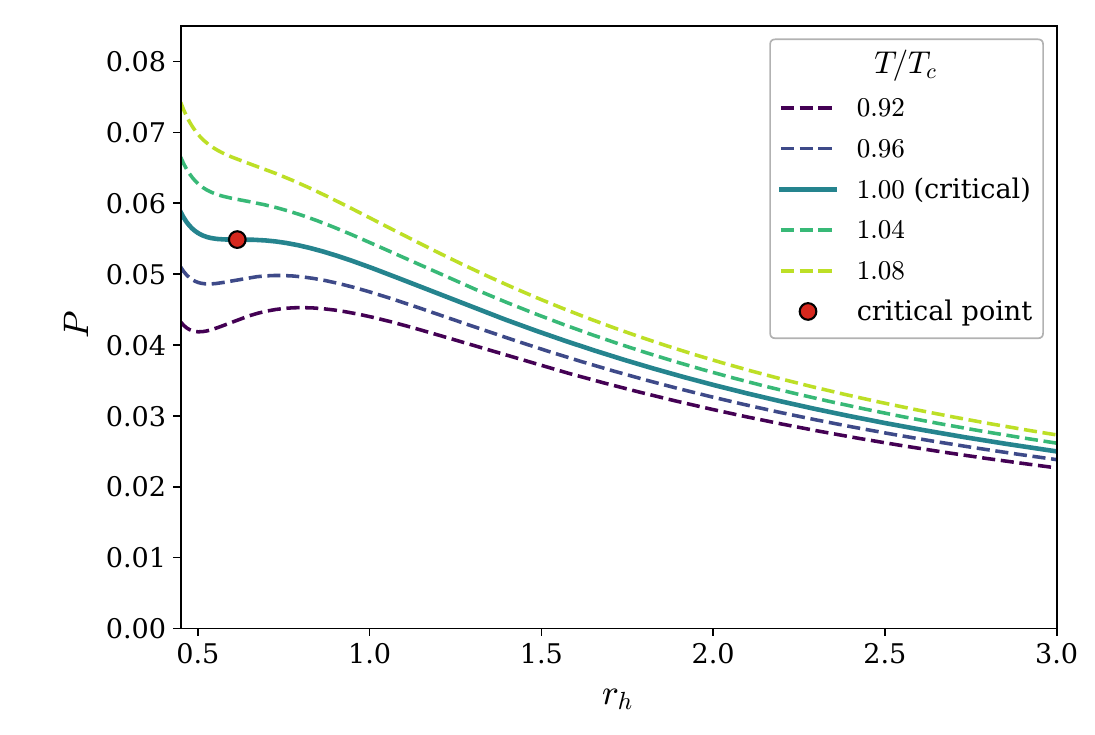} &
\gpanel[width=0.47\textwidth]{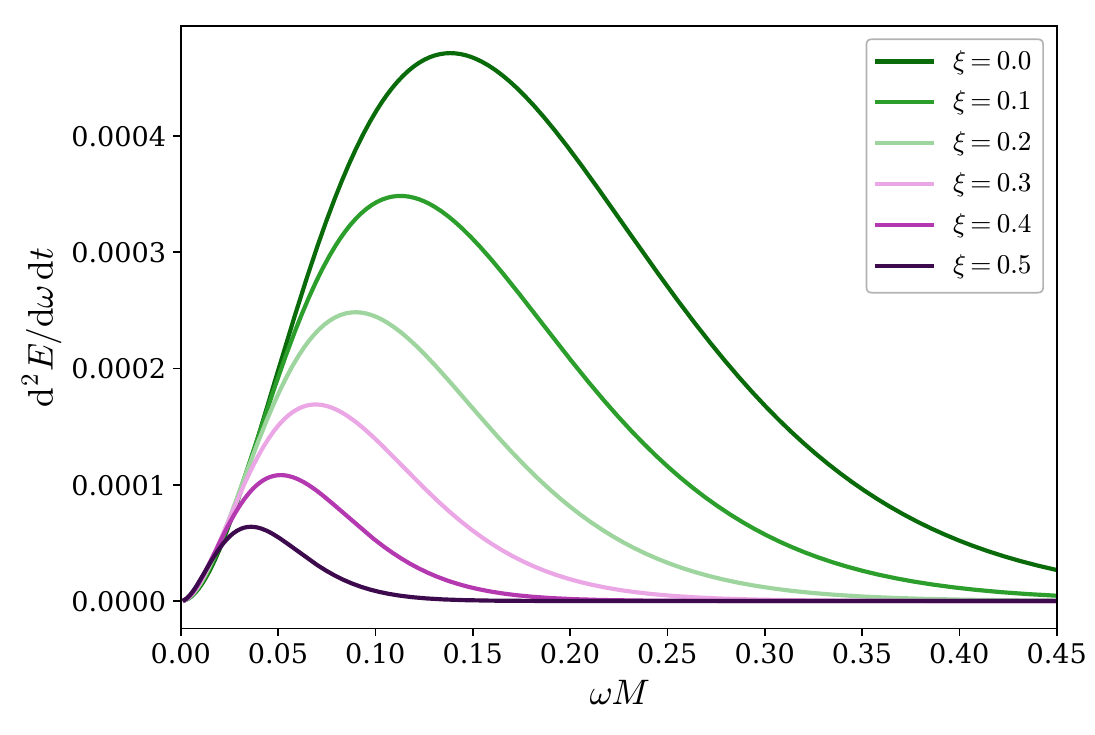}\\[2pt]
(a) & (b)
\end{gridfig}
\caption{\label{fig:pvem}(a) $P$--$r_{h}$ isotherms at $\xi=0.1$ ($Q=0.3$, $\beta=5$, $\lambda=0.1$) for $T/T_{c}=0.92,0.96,1.00,1.04,1.08$; the critical isotherm (solid) shows the inflection with horizontal tangent and the marker is the critical point of Table~\ref{tab:crit}, with oscillating isotherms below $T_{c}$ and monotonic ones above. (b) Energy emission rate $\dd^{2}E/\dd\omega\,\dd t$ from Eq.~\eqref{emission} for $Q=0.4$, $\beta=1$, $\lambda=0.1$ {\color{black} in the asymptotically flat limit $\Lambda\to0$}, colored by $\xi\in[0,0.5]$; the peak falls and reddens as the defect strengthens.}
\end{figure}

The same horizon data fix the thermal radiation the \BH{} emits. In the geometric-optics regime the energy emission rate per unit frequency takes the Stefan--Boltzmann form~\cite{WeiLiu2013,Decanini2011}
\begin{equation}
\frac{\dd^{2}E}{\dd\omega\,\dd t}=\frac{\sigma_{\rm lim}}{2\pi^{2}}\,\frac{\omega^{3}}{e^{\omega/T_{H}}-1},\qquad \sigma_{\rm lim}=\pi R_{\rm sh}^{2},
\label{emission}
\end{equation}
where $\sigma_{\rm lim}$ is the high-frequency absorption cross section, fixed by the shadow radius of Sec.~\ref{isec3}, and $T_{H}$ the temperature above. The defect pulls the two inputs in opposite directions: a larger shadow raises $\sigma_{\rm lim}$, while the lower temperature softens the Planck factor. Figure~\ref{fig:pvem}(b) shows the spectrum {\color{black} in the asymptotically flat limit $\Lambda\to0$}. The temperature drop wins. {\color{black} As $\xi$ grows from $0$ to $0.5$ the temperature falls from $T_{H}M=0.049$ to $0.013$, and although the capture cross section more than doubles, the peak emission rate drops by roughly a factor of seven and shifts to lower frequency.} A stronger defect thus makes the \BH{} cooler, dimmer, and redder, so the thermal channel reinforces the ringdown and greybody trends. The sparsity of the flux rises in step, since a colder horizon emits quanta further apart relative to their wavelength.

\section{Conclusions}\label{isec8}
We assembled a static, spherically symmetric \BI{} \AdS{} \BH{} carrying a topological defect and sitting in \PFDM{}, then tracked how its four sectors imprint on imaging, timing, ringdown, and thermodynamics. The limits of Sec.~\ref{isec2}, all confirmed symbolically, anchor the solution.

The defect strength $\xi$ proved the most visible of the four parameters. It pushes the horizon, photon sphere, and shadow outward together (Table~\ref{tab:horizon}), drives the ISCO from $4.94M$ to $9.76M$ while holding the radiative efficiency near $5\%$ (Table~\ref{tab:isco}), softens and lowers the scalar ringdown (Table~\ref{tab:qnm}), and moves the heat-capacity divergence separating the stable and unstable branches. The twin-peak QPO fit holds the defect away from zero at the few-$\sigma$ level for GRO~J1655--40 and GRS~1915+105, with the caveat that a static metric underpredicts the masses of rapidly spinning microquasars. The \BI{} scale regularizes the charged core, the halo intensity deepens the potential well logarithmically, and the \AdS{} pressure organizes a \VdW{}-type phase structure whose critical point drifts with all three remaining parameters. Taken together, the four sectors leave distinct signatures: the inner-disk radius tracks the defect almost independently of the efficiency, while the ringdown and thermodynamic channels respond to the charge and pressure in ways imaging alone would miss. {\color{black} We stress that these are statements of parameter sensitivity, not of demonstrated observational separability: the analysis varies one sector at a time, so it shows which observable responds most strongly to which parameter, whereas disentangling the four from data would need a simultaneous multi-parameter fit with the degeneracies mapped. In the same spirit, the defect values returned by the QPO fit are model-dependent indications under the relativistic-precession identification rather than calibrated constraints.}

Rotation is the most pressing of the follow-up problems this leaves open. Our fit underpredicts the microquasar masses precisely because the metric is static, so a rotating extension through the Newman--Janis algorithm, or a direct stationary solution, would lower the innermost stable orbit, raise the attainable frequencies, and let the same data constrain defect and spin jointly. The ringdown beyond the fundamental mode is a second direction: the WKB spectrum computed here can be pushed to the overtones and to electromagnetic and gravitational perturbations, then confronted with time-domain integration to test where the WKB estimate degrades for the \AdS{} barrier. A third is the thermal sector, where a greybody-bound analysis together with the sparsity of the Hawking flux would complete the emission picture. On the observational side, the angular-diameter table invites a joint fit of shadow size and twin-peak frequencies for Sgr~A$^{\ast}$ and stellar-mass sources, since both reduce to one defect parameter. The extended phase space analysis can also be carried to the microstructure and holographic readings of the critical point, and to generalized-entropy corrections. We intend to pursue the rotating generalization and the joint imaging--timing fit first, since those bear most directly on data already in hand.

\section*{acknowledgments}

{\color{black} The authors thank the Editor-in-Chief, the Handling Editor, and the anonymous Referees for the careful reading of the manuscript and for the constructive comments, which improved the presentation of this work.} F.A.\ acknowledges the Inter University Centre for Astronomy and Astrophysics (IUCAA), Pune, India, for the granting of visiting associateship. The authors acknowledge Eastern Mediterranean University, T\"UB{\.I}TAK, ANKOS, and SCOAP3. \.I.S. is grateful for the networking support of COST Actions CA18108, CA22113, CA21106, CA23130, CA23115, and CA21136.

{\small\linespread{0.86}\selectfont
\setlength{\bibsep}{0pt plus 0.1pt}
\setlength{\parskip}{0pt}
\bibliographystyle{ws-ijgmmp}
\bibliography{ref6}
}

\end{document}